\documentclass[aps,prx,twocolumn,floatfix,superscriptaddress]{revtex4-2}
\usepackage{amsmath}  
\usepackage{amsfonts} 
\usepackage{graphicx} 
\usepackage{epstopdf}
\usepackage{gensymb}
\usepackage{color}
\usepackage{soul}
\newcommand{\edits}[1]{{\color{black}#1}}

\begin{document}

\title{{Mechanical properties} of acoustically levitated granular rafts} 
\author{Melody X. Lim}
\email{mxlim@uchicago.edu}
\affiliation{James Franck Institute, The University of Chicago, Chicago, Illinois 60637, USA}
\affiliation{Department of Physics, The University of Chicago, Chicago, Illinois 60637, USA}
\author{Bryan VanSaders}
\affiliation{James Franck Institute, The University of Chicago, Chicago, Illinois 60637, USA}
\author{Anton Souslov}
\affiliation{James Franck Institute, The University of Chicago, Chicago, Illinois 60637, USA}
\affiliation{Department of Physics, University of Bath, Bath BA2 7AY, United Kingdom}
\author{Heinrich M. Jaeger}
\affiliation{James Franck Institute, The University of Chicago, Chicago, Illinois 60637, USA}
\affiliation{Department of Physics, The University of Chicago, Chicago, Illinois 60637, USA}

\begin{abstract}
We investigate a model system for the rotational dynamics of inertial many-particle clustering, in which sub-millimeter objects are acoustically levitated in air. Driven by scattered sound, levitated grains self-assemble into a monolayer of particles, forming mesoscopic granular rafts with both an acoustic binding energy and a bending rigidity. Detuning the acoustic trap can give rise to stochastic forces and torques that impart angular momentum to levitated objects. As the angular momentum of a quasi-two-dimensional granular raft is increased, the raft deforms from a disk to an ellipse, eventually pinching off into multiple separate rafts, in a mechanism that resembles the break-up of a liquid drop. We extract the raft effective surface tension and {elastic modulus}, and show that {non-pairwise} acoustic forces give rise to effective elastic moduli that scale with the raft size. We also show that the raft size controls the microstructural basis of plastic deformation, resulting in a transition from fracture to ductile failure.
\end{abstract}
\maketitle
\section{Introduction}

The dynamics of rapid rotation underpin a wide range of physical systems, from rotating black holes~\cite{smarr1973mass,genzel2003near,mckinney2004measurement}, to the shapes of spinning self-gravitating asteroids~\cite{walsh2008rotational,rozitis2014cohesive,barnouin2019shape}, the cooling of optically trapped microparticles~\cite{arita2013laser}, and the spin and stability of atomic nuclei~\cite{cohen1974equilibrium,pomorski2003nuclear,schunck2007nuclear, arabgol2019observation}. 
To probe the stability and modes of deformation of such systems, rotating liquid droplets are often used as models, where surface tension mimics attractive forces that bind the material and compete with the outward pressure exerted by the
rotation~\cite{cohen1974equilibrium,smarr1973mass,hill2008nonaxisymmetric,chandrasekhar1965stability,brown1980shape,pomorski2003nuclear,schunck2007nuclear}. The small size of molecules implies that liquid droplets can only represent the elastic limit where the number of constituent components is very large, and internal structure is treated as a continuum.
Deviations from elastic behavior emerge in the mesoscopic regime as the surface to volume ratio becomes large. For example, measurements of thin films and metallic nanopillars have found elastic moduli~\cite{chen2006size,agrawal2008elasticity}, dielectric constants~\cite{yang2012size}, and melting temperatures~\cite{koga2004size} that depend on the system size. 


Observing the effects of internal structure and investigating emergent properties as a function of the number of constituent particles in the mesoscopic regime is possible by using colloids~\cite{anderson2002insights,poon2004colloids,manoharan2015colloidal} or the micron-sized particles in dusty plasmas~\cite{rubin2006kinetic,knapek2007recrystallization,sutterlin2009dynamics} as `model atoms’, but generating rapid rotations is difficult. 
For exploring the mesoscale dynamics in a rapidly rotating, inertial, many-particle system we here introduce acoustically levitated granular rafts: close-packed monolayers of sub-millimeter particles freely floating in air.  

In these rafts, tunable attractive forces generate short-ranged cohesion, providing in-plane elastic properties as well as out-of-plane bending stiffness. 
As the rotation rate increases beyond the point where inertia outweighs this cohesion, sufficiently large rafts undergo a shape transition strikingly similar to liquid drops. 
We can image such granular rafts on the particle scale at high temporal resolution, allowing us to measure microstructural properties and dynamics during inertial driving. 
To explore the emergent physics of this granular system, we here focus on mesoscale rafts comprised of 10 to 200 particles.

\begin{figure*}
\centering
\includegraphics[width = 2\columnwidth]{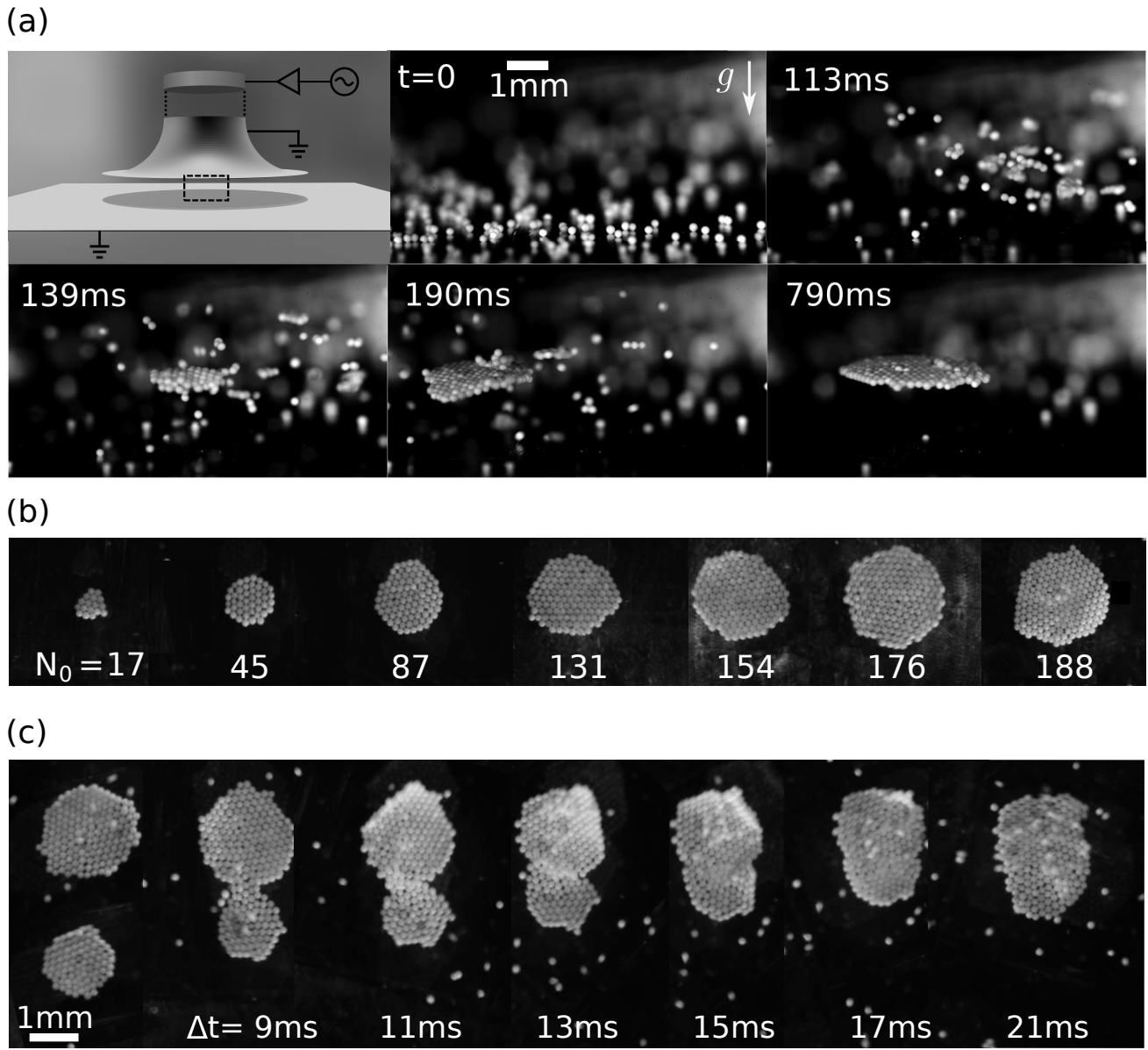}
\caption{Self-assembly of granular rafts by acoustic levitation. (a) 3D drawing of setup and sequence of images (side view) showing the self-assembly of a granular raft from its constituent particles. Piezoelectric elements (cylinder) are attached to an aluminium horn to generate ultrasound (only base is shown). The grounded aluminium horn is spaced over a (grounded) indium tin oxide glass slide. The dashed black rectangle indicates the field of view in the subsequent still images. At~$t=0$, a standing wave is established between the transducer (above the top of the image) and reflector surface, which is covered with loose particles. These particles are picked up from the reflector surface by the primary acoustic force, and initially form small clusters, which travel in the underdamped acoustic environment until they coalesce to form a monolayer. (b) Self-assembled rafts composed of varying numbers of particles~$N_0$, viewed from below. (c) Sequence of images from below, showing two rafts approaching each other and merging into a larger raft. Variations in brightness correspond to local curling of the raft out of plane.}
\label{fig:assemble}
\end{figure*}

Our setup consists of a cylindrical ultrasound transducer (Langevin horn) and a reflector, between which we generate a standing sound wave with a single pressure node along the vertical direction (Fig.~\ref{fig:assemble}(a))~\cite{lim2019cluster,lim2019edges}.
While strong sound pressure enables the levitation of solid particles (primary acoustic force), sound scattering between the particles generates attractive interactions among them (secondary acoustic force)~\cite{silva2014acoustic, sepehrirahnama2015numerical}. 
Particles are levitated in air, generating an underdamped environment in which levitated particles collide and self-assemble into a raft, weakly confined to the horizontal plane of the sound pressure node (Fig.~\ref{fig:assemble}(a), Supplementary Movie 1). 
Such rafts form roughly circular monolayers comprised of varying numbers of constituent particles (Fig.~\ref{fig:assemble}(b)).

Driving the cavity slightly above resonance produces stochastic, non-conservative forces, including a torque (along the vertical direction) that imparts angular momentum to the levitated rafts and spins them up. 
Unlike other strategies for activating underdamped matter~\cite{tsai2005chiral,briand2016crystallization,scholz2018inertial}, the resulting motion of the granular rafts is entirely substrate free.
At low angular velocities, the rafts retain their close-packed internal structure (visible in Fig.~\ref{fig:assemble}(b)).
As their rotation speeds up, the rafts undergo deformation via internal rearrangements until they eventually break apart into smaller fragments. 
Weak radial confinement within the nodal plane brings these pieces back together and they merge by forming a bridge that grows with time, eventually coalescing  back into a single circular raft (Fig.~\ref{fig:assemble}(c), Supplementary Movie 2).
This process is similar to the surface-tension driven coalescence of a pair of liquid drops~\cite{thoroddsen2005coalescence,aarts2005hydrodynamics}.
Once merged, the (again) circular drop can then be spun up by the acoustic torque, repeating the cycle of spin-up-to-failure.

The spin-up during such cycles provides conditions in which the gyrostatic pressure slowly increases, driving micro-structural changes as well as overall shape deformations. {These shape deformations reveal the consequences of the binding potential, here induced by acoustic scattering, on measurable mechanical properties of the granular rafts. 

Specifically, we use the rotation-induced shape changes to track how the emergent mechanical properties depend on raft size. We find that the effective surface tension and the effective {elastic modulus} grow roughly linearly with the number of particles, i.e. both are extensive quantities. We show that this is a direct consequence of acoustically-induced binding energies. This extensivity demonstrates the presence of non-pairwise acoustic forces between particles. We conclude that the properties of levitated granular rafts challenge existing frameworks for the calculation of acoustic forces between multiple particles, which assume superposition of pairwise interactions and, over the same range in raft sizes we explore, would predict saturation at some elastic value. We further show that these rafts also display size-dependent microstructural deformations. }


\section{Experimental Setup and Methods}

\subsection*{Experiments}

Our acoustic trap consists of an acoustic resonant cavity, driven on one side by a commercial ultrasound transducer (Hesentec HS-4SH-3840). 
An aluminum horn was bolted onto the transducer to maximize the strength of the pressure field, as detailed in Ref.~\cite{lim2019cluster,lim2019edges} (Fig.~\ref{fig:assemble}(a), first panel). 
The base of the horn (diameter 38.1mm) was painted black to better image the particles from below. 
The transducer was driven by applying a sinusoidal signal of peak-to-peak voltage~$V_{pp}$ (100 - 400V) and frequency~$f$ close to {the resonance frequency of the horn}~$f_0 = 45.65$kHz, produced by a function generator (BK Precision 4052) connected to a high-voltage amplifier (A-301 HV amplifier, AA Lab Systems). 
The transducer-reflector distance was adjusted via a translation stage to~$\lambda_0/2 = 3.8$mm, establishing a single pressure node within the acoustic trap. 
Stable levitation is possible across a range of a few tens of Hz to either side of the resonant frequency. 
In order to reduce the effects of air currents, the entire setup was enclosed in a transparent acrylic box, with side-walls far from the experimental area of interest ($l\times w\times h= 24\times 12 \times 18$ in$^3$). 

We used polyethylene spherical particles (Cospheric, material density~$\rho = 1,000\:\mathrm{kg}\: \mathrm{m}^{-3}$, diameter~$d=180-200\: \mu \mathrm{m}$). 
The particles were stored and all experiments were performed in a humidity- and temperature-controlled environment (40-50\% relative humidity, 22-24$\degree$C). 
The reflector was comprised of a grounded indium tin oxide (ITO) coated glass slide (thickness 1.1mm) secured to the top of an acrylic sheet (thickness 6.35mm). 
To mitigate tribocharging, both the reflector and the horn were grounded.
The setup was cleaned with compressed air, ethanol and de-ionized water before each experiment. 
We neutralized any charges that remained on the reflector with an anti-static device (Zerostat 3, Milty).  
For each experimental run, particles were scattered onto the reflector from a spatula or, in some cases, inserted with a tweezer. 
Video was recorded with a high-speed camera (Vision Research Phantom v12) at 3,000 frames per second. 

\subsection*{Lattice Boltzmann Simulations}

\edits{Unlike other forms of fluctuation induced forces, such as critical Casimir forces~\cite{hertlein2008direct}, acoustic forces result from the inclusion of a rigid object in a highly structured (single-frequency) field. In order to perform \textit{ab initio} simulations of the sound field inside the acoustic cavity and its coupling with the levitated particles, we employ the Lattice Boltzmann method (LBM). These simulations take} into account the full extent of the fluid-structure interactions~\cite{yu2003visc, bauer2020walberla}. 
This approach naturally includes the effects of viscous dissipation, momentum transfer due to multiple scattering events, and anisotropy in the shape of the levitated objects. 

LBM simulations of the acoustic cavity were carried out within the \texttt{waLBerla} framework \cite{bauer2020walberla}.
A single relaxation time scheme with a viscosity matching that of air was used \cite{yu2003visc}.
To compute inter-particle forces, a simplified simulation geometry with plane wave acoustic input and periodic domain boundaries was used.
The ultrasonic horn was represented as a bounce-back boundary condition with time-dependent velocity, and the reflector as a stationary no-slip boundary.

We used the \texttt{PE} functionality of the \texttt{waLBerla} framework to simulate the interaction of particles with the acoustic field~\cite{gotz2010pe}.
Hydrodynamic forces between the particles and fluid were handled with the partially-saturated cells method \cite{owen2011efficient}, which was found to be more stable than other momentum-exchange methods under acoustic conditions.
We found that a local cell size of~$D/15$, where~$D$ is the particle diameter, was sufficient for accurate force calculations.

\begin{figure}
\centering
\includegraphics[width = 1\columnwidth]{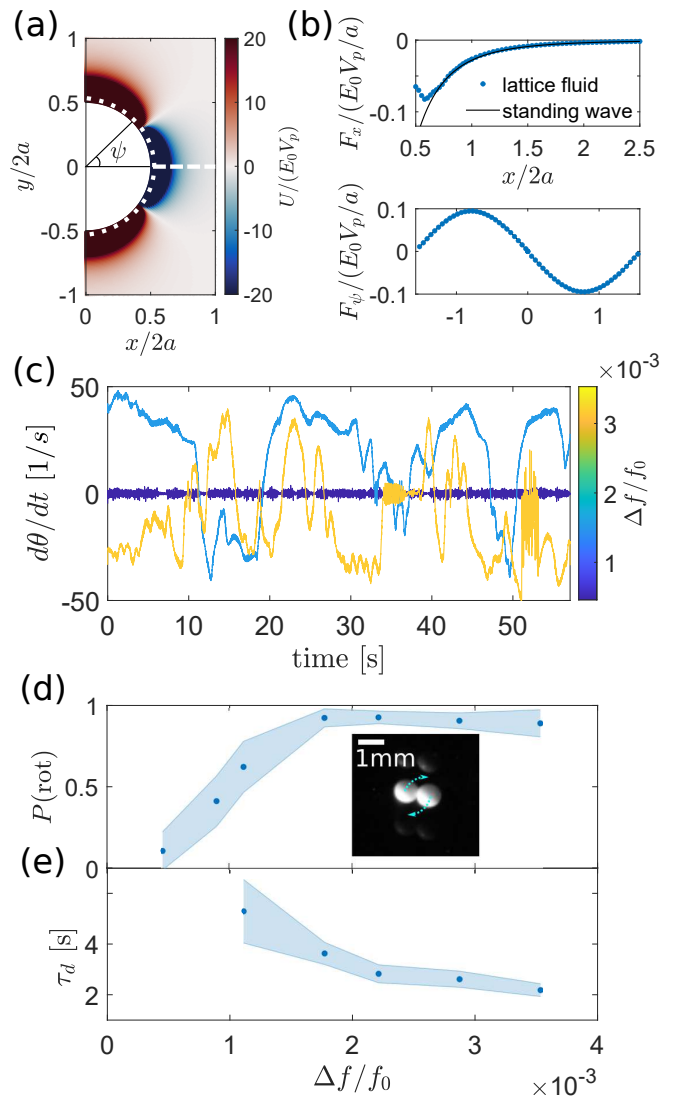}
\caption{Secondary acoustic forces drive self-assembly and rotation of levitated granular rafts. (a) Secondary acoustic potential due to a particle with radius~$a$, computed using a scattering expansion, normalized by the energy density of the cavity~$E_0$, and the particle volume~$V_p$. (b) Secondary acoustic force between a pair of particles, computed using a scattering expansion (black solid line), and using a lattice simulation (blue markers). (top) Horizontal force~$F_x$ between two particles as a function of distance~$x$ between their surfaces (dashed white horizontal line in panel (a)). (bottom) Angular force~$F_\psi$ on a particle displaced out of the nodal plane with angle~$\psi$ ($x/2a=0.1$, dotted circular contour in (a)). (c) Example traces of rotation rate~$d\theta/dt$ as a function of time, for a 2-particle dimer {consisting of a pair of 700-810$\mu$m polyethylene particles}, at different detuning parameters~$\Delta f/f_0$. (d) Probability of dimer rotation~$P(\mathrm{rot})$ as a function of~$\Delta f/f_0$. Shaded areas indicate the standard error. (e) Plot of the decorrelation time~$\tau_d$ for the rotation rate of a pair of particles, as a function of~$\Delta f/f_0$. Shaded areas indicate the standard error. Data not plotted for the smallest detuning parameters, for which there is no significant change in the rotation over time. }
\label{fig:setup}
\end{figure}

In Fig.~\ref{fig:setup}(b) we compare the results produced by the analytical approximation~\cite{silva2014acoustic} and our LBM simulations for the secondary acoustic force due to scattering between two particles of radius~$a\ll \lambda_0$ (Rayleigh limit) and volume~$V_p$, in an imposed standing wave with acoustic energy density~$E_0$.
The data show close agreement, except very close to the particle surface, where the far-field scattering approximations break down and viscous effects become increasingly important~\cite{settnes2012forces}. 
Spherically symmetric particles in the same horizontal plane experience an azimuthally symmetric acoustic potential well (Fig.~\ref{fig:setup}(a)). 
As a result, the secondary acoustic force along the horizontal direction~$x$ is attractive, driving the particles into direct contact (Fig.~\ref{fig:setup}(b)). 
For particles approaching at some angle~$\phi$ with respect to the horizontal, the interaction is more complex due to the quadrupole-like secondary acoustic potential (Fig.~\ref{fig:setup}(a)).
The result is a restoring force that brings particles back into the plane (Fig.~\ref{fig:setup}(b)). 
In both cases the secondary acoustic force is short-ranged, acting on lengthscales of the particle radius.
This gives rise to an effective cohesion and bending rigidity of the raft, while at the same time stabilizing monolayer formation by penalizing particles for stacking into multiple layers. 
In contrast, the primary acoustic force, which sets up an acoustic potential well centered around the nodal plane of the standing wave in the cavity, has a characteristic length scale on the order of the wavelength of sound. 
Together, this means that the levitated granular rafts behave as effectively two-dimensional membranes that are weakly confined in three dimensions.

\subsection*{Raft Rotation}

In addition to conservative acoustic forces, which assemble and stabilize the rafts, non-conservative forces can be generated if the transducer is driven with a frequency slightly larger than the resonance condition for levitation.
These non-conservative forces can manifest as of out-of-plane, velocity-dependent forces, which originate from a phase lag between the motion of levitated objects in the cavity and the response of the cavity mode~\cite{rudnick1990oscillational}, and lead to vertical height fluctuations.
In prior work we showed that such fluctuations can drive cluster rearrangements~\cite{lim2019cluster}, or actuate modes of deformation within a cluster held together by secondary acoustic forces~\cite{lim2019edges}.

Our focus here is on fluctuating torques caused by off-resonance driving, where angular momentum is transferred to levitated objects, such that they spin around an axis perpendicular to the nodal plane. 
{These spontaneous, fluctuating torques are frequently observed in levitated objects in air. 
Candidates to explain their origin include streaming flows in the acoustic cavity~\cite{trinh1994experimental}, velocity-dependent instabilities~\cite{rudnick1990oscillational}, and radial potential gradients in the levitation plane~\cite{baer2011analysis}. 
Although a full explanation of the origin of this acoustic torque is outside of the scope of the present study, here we show that the statistics of this oscillational instability can be controlled by the transducer frequency detuning~$\Delta f/f_0$, such that they can be used to provide rotational driving to a levitated granular clusters.}

To demonstrate and characterize this momentum transfer as a function of frequency detuning~$\Delta f/f_0$, we measure the in-plane angular rotation rate for a cluster consisting of two (700-810$\mu$m polyethylene) particles held together by the secondary acoustic force (data in Fig.~\ref{fig:setup}(c), see inset to Fig.~\ref{fig:setup}(d) for an image).
{We chose to use a pair of larger spheres as a minimal model for these rotational measurements, such that the longest dimension of the pair together was approximately equal to the diameter of the largest rafts. 
Results are qualitatively similar for pairs of smaller spheres, rigid rods, and the rafts, with differences in the maximum possible rotation rate and acceleration due to size-dependent viscous drag.}

For the smallest~$\Delta f/f_0$, the cluster does not complete full rotations, but simply rocks back and forth, occasionally stopping. 
As~$\Delta f/f_0$ is increased, the cluster spins up to high angular speed, where both rotation rate and rotation direction vary stochastically when tracked over tens of seconds. 
However, this also includes stretches where the rotation rate increases roughly linearly with time, such that the imparted torque is nearly constant. 
Such stretches are used for our measurements on the rafts, reported below, which involve time intervals of typically less than one second.
With increasing detuning, the probability of being in some continuously rotating state quickly approaches unity (Fig.~\ref{fig:setup}(d)), while the typical lifetime of the state, measured by the time for the auto-correlation of the angular speed to decay to half its value, never drops below two seconds (Fig.~\ref{fig:setup}(e)). 

\section{Results and Discussion}

\begin{figure*}
\centering
\includegraphics[width = 1.6\columnwidth]{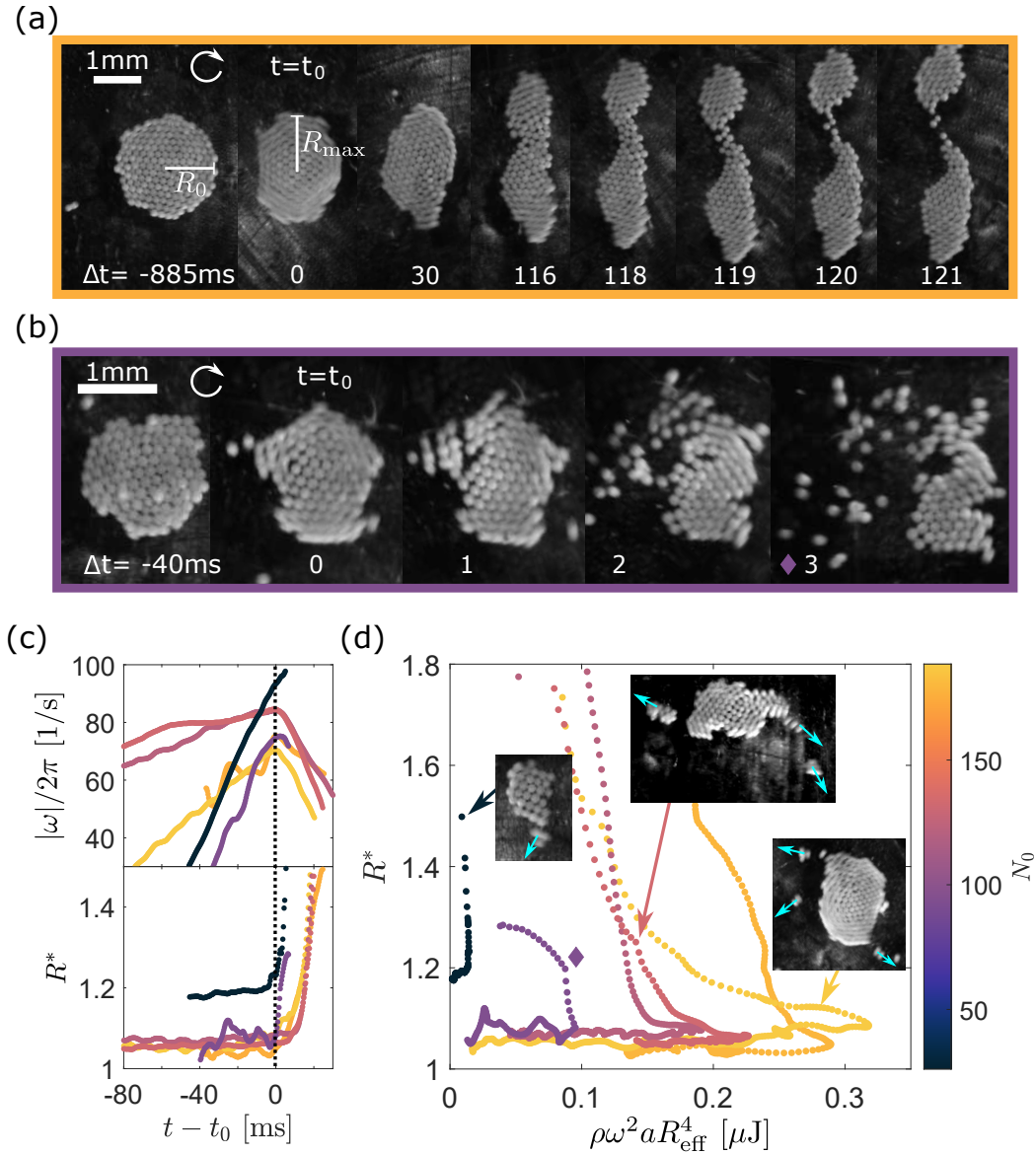}
\caption{Levitated granular rafts exhibit emergent liquid-like behavior. 
(a) Sequence of images from below, showing the deformation of an initially circular raft that rotates in the clockwise direction. As the raft gains angular momentum, it elongates into an ellipse (at~$t=t_0$, where the kinetic energy~{$\rho\omega^2 a R_\mathrm{eff}^4$} is at its maximal value), then splits into two smaller rafts. The corresponding traces are shown in orange in (c) and (d) (second largest raft) (b) Sequence of images from below, showing the deformation of a smaller raft, at first also by elongating (at~$t=t_0$), but then by shedding particles. The corresponding traces are shown in purple in (c) and (d) (second smallest raft), with the diamond marking the image at time~$t=3$ms in (b).  (c) Example time-series of the absolute value of the number of rotations per second~$|\omega|/2\pi$ (top), and the dimensionless shape parameter~$R^*$ of the levitated raft (bottom), as a function of the number of particles in the raft~$N_0$, see part (d) for color key. Before~$t_0$, the rafts spin faster without deforming; after~$t_0$, the rafts change shape, shed weakly bound particles, and eventually split into multiple pieces. Raw data was filtered using a moving-average filter, with a width of 20 data points (7ms) to obtain the data in (c) and (d).  (d) Evolution of the dimensionless shape parameter~$R^*$, as a function of the kinetic rotational energy of the drop~$\rho\omega^2 a R_\mathrm{eff}^4$, for rafts with several different~$N_0$.  Inset images show the rafts at various points in the spinning process. After deformation into an ellipse, rafts can shed angular momentum by losing small clusters of weakly bound particles.}
\label{fig:drops}
\end{figure*}

As the initially roughly circular rafts spin up, their rotational kinetic energy eventually becomes comparable to the particle binding energy and their shape begins to deform into ellipses (Fig.~\ref{fig:drops}, Supplementary Movie 3). 
This plastic deformation process (see Appendix D for Voronoi diagrams of the raft interior) continues until eventual break-up.
For sufficiently large rafts, plastic deformation is localized mainly to a small ``neck" region that continues to extend as the raft rotates, eventually pinching off into two or more raft pieces  (Fig.~\ref{fig:drops}(a)). 
In contrast, for smaller rafts we observe a different mode of shape change (Fig.~\ref{fig:drops}(b)): after the rafts initially elongate into ellipses, they tend to continue to deform by shedding particles from their perimeter rather than via plastic deformation. 


We now focus on the behavior well before break-up, where the rafts first begin to deviate from their circular shape and to elongate into ellipses. 
For a raft of total area~$A$, which may change as the raft gains rotational kinetic energy, we measure deviations from circularity by the dimensionless parameter~$R^*= R_\mathrm{max}/R_\mathrm{eff}$.
Here~$R_\mathrm{max}$ is the semi-major axis of the raft at any given time, and~$R_\mathrm{eff}=\sqrt{A/\pi}$ is the effective radius of the raft’s (time-varying) area $A$.

Comparing time-traces of the rafts’ rotation rate~$\omega$ and shape parameter~$R^*$ (Fig.~\ref{fig:drops}(c)) reveals that the shape evolution is divided into two regimes.
At first,~$\omega$ increases nearly linearly with time, while~$R^*$ remains close to its initial value, indicating constant torque and angular acceleration without shape change -- rotational kinetic energy is diverted into uniform stretching of the acoustic ``bonds" between the constituent particles of the raft. In this regime, the {raft perimeter} is elastically stretched by increasing the interparticle spacing, without moving particles from the interior to the raft {perimeter}.    

Once a maximum spinning speed has been reached, a point in time we label as $t_0$ in Fig.~\ref{fig:drops}(c), the spinning speed~$\omega$ decreases, and the shape parameter~$R^*$ increases sharply -- the rafts become elliptical, and increase the total length of their perimeter by introducing particles from the interior to the raft surface. 
We use these differences in the microstructural evolution of the rafts to distinguish the two raft regimes as being dominated by elasticity before~$t_0$, and being dominated by surface tension at and after~$t_0$ {(where ``surface" here refers to the outer perimeter of the raft)}, in accordance with the nomenclature used in studies of the surface stresses and energies of thin films~\cite{cammarata1994surface, haiss2001surface}.

In Fig.~\ref{fig:drops}(d) we plot the evolution of~$R^*$ as a function of the rotational kinetic energy of the rafts $E_{rot}$.
For our circular monolayer rafts~$ E_\mathrm{rot}\sim\rho \omega^2 a R_\mathrm{eff}^4$, where we have treated the rafts as discs with the thickness of one particle (diameter~$2a$ and material density~$\rho$), and where we have omitted numerical prefactors of order unity. 
After reaching a maximum $E_\mathrm{rot}$, the rafts lose rotational kinetic energy, visible as a change in curvature in the traces. 
After this point,~$R^*$ grows as the rotational kinetic energy decreases: further increases to the angular momentum of the raft serve to increase its moment of inertia, increasing its surface area as it elongates. 
While the raft continues to lengthen, ejection of weakly bound particles, which carry away some of the angular momentum, is reflected in sharp curvature changes of the traces in Fig.~\ref{fig:drops}(d).
The insets give examples.  

The general shape of the traces in Fig.~\ref{fig:drops}(d) exhibits striking similarity with what is found for rotating droplets of molecular liquids~\cite{chandrasekhar1965stability,brown1980shape, trinh1988acoustic,tian1995new}. 
For liquid drops, as with our granular rafts, shape is governed by the competition between rotational kinetic energy, $E_\mathrm{rot}$, which acts to elongate the drop, and interfacial energy, $E_\mathrm{int}$, which penalizes increases to {the drop-air interfacial area (for rafts this is the perimeter area, i.e. the product of the perimeter and~$2a$). 
Air drag can be neglected provided that the raft undergoes rigid body rotation~\cite{brown1980shape,wang1994bifurcation} (see Appendix E for measurements of the non-affine motion of particles in the raft, and Appendix F for an estimate of the magnitude of drag on the edge of the raft, which we show to be much smaller than the secondary acoustic forces binding particles to the raft edge). 
Although these aerodynamic forces are small compared to the in-plane forces holding raft particles together, these small forces can excite resonant out-of-plane bending similar to what is observed for thin, flexible disks that are spinning rapidly (see Appendix G for measurements and a qualitative discussion). }

In liquids, as the spinning speed increases under constant torque, the droplets stay axisymmetric, and the energy ratio $\Sigma = E_\mathrm{rot}/E_\mathrm{int}$ is simply a function of the spinning speed~$\omega$ until a value~$\Sigma_\mathrm{max}$ is reached.
Beyond this value there is a bifurcation away from the axisymmetric shape of the drop, such that the drop elongates along one axis, decreasing its rotation rate~\cite{chandrasekhar1965stability,brown1980shape}. 
Chandrasekhar~\cite{chandrasekhar1965stability} calculated the maximum ratio value~$\Sigma_\mathrm{max}\approx 4$ for 3D droplets, a value that is found to increase to 12 for droplets in two dimensions~\cite{lewis1987stability}. 

As Fig. 3(d) implies, the ratio $\Sigma$ for our rafts similarly increases up to some maximum value $\Sigma_\mathrm{max}$, beyond which there is a significant change away from the initially circular shape and the rotational kinetic energy decreases.
Not directly apparent from the data in Fig. 3(d) (but discussed further below) is a slight isotropic expansion of the raft that occurs without significant change in~$R^*$ up to the point that $\Sigma_\mathrm{max}$ is reached. 
Here we note that this dilation serves to drive the particles toward a state where they can begin to plastically flow. 
For such flow, the important quantity is the energy associated with the creation of new interfaces. 

Thus, by identifying the maximum~$E_\mathrm{rot}$ before shape change occurs in data such as Fig.~\ref{fig:drops}(d), we can extract the interfacial energy~$E_\mathrm{int} = E_\mathrm{rot,max}/\Sigma_\mathrm{max}$ of the rafts, in analogy to liquid droplets.
{Using~$E_\mathrm{int} = \gamma a R_\mathrm{eff}$, we define an effective surface tension $\gamma$: 

\begin{align}
    \gamma = \frac{E_\mathrm{rot,max}}{2a\Sigma_\mathrm{max}R_\mathrm{eff}} \, .
    \label{eq:gamma}
\end{align}}

\begin{figure*}
\centering
\includegraphics[width = 2\columnwidth]{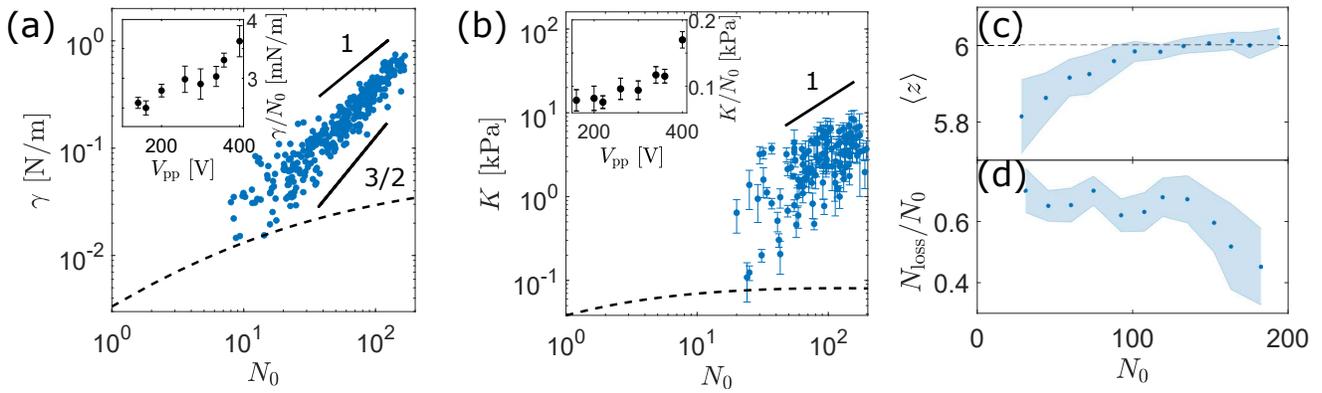}
\caption{Contactless measurement of effective surface tension, effective {elastic modulus}, and microstructural properties. (a) Effective surface tension~$\gamma$ as a function of the initial number of particles in a raft~$N_0$. Dashed line displays the prediction for~$\gamma$ obtained from integrating the acoustic force (Fig.~\ref{fig:setup}(a)) over the raft body (within a multiplicative prefactor, see Appendix A). Data shown for transducer peak-to-peak driving voltage~$V_\mathrm{pp}=$300V. (inset) The slope of~$\gamma$ with respect to~$N_0$, measured for different~$V_\mathrm{pp}$. Error bars indicate standard error. (b) Effective {elastic modulus}~$K$, {defined in analogy to the bulk modulus of a three-dimensional material}, as a function of the initial number of particles in a raft~$N_0$. Dashed line displays the prediction for~$K$ obtained from integrating the acoustic force (Fig.~\ref{fig:setup}(a)) over the raft body, with a multiplicative prefactor (see Appendix A). Data shown on a log-log plot for transducer peak-to-peak driving voltage~$V_\mathrm{pp}=$ 300V. Error bars indicate error due to fitting the data. (inset) The slope of~$K$ with respect to~$N_0$, measured for different~$V_\mathrm{pp}$. Error bars indicate standard error. (c) Average coordination number of particles in the raft interior~$\langle z\rangle$, as a function of~$N_0$. Dotted line indicates the coordination number for close-packed configurations in two dimensions. (d) Lost number of particles up to maximal plastic deformation,~$N_\mathrm{loss}$, normalised by~$N_0$, as a function of~$N_0$. Shaded regions indicate the standard error. }
\label{fig:st}
\end{figure*}

Our data, plotted in Fig.~\ref{fig:st}(a), reveals that the effective surface tension of the granular rafts is extensive, and scales as a power law with the number of constituent particles,~$N_0$. This scaling of the effective surface tension does not depend on the numerical value for~$\Sigma_\mathrm{max}$, which we here take to be 12, in accordance with Ref.~\cite{lewis1987stability}.
Within the scatter of the data, there is no consistent trend with the detuning parameter~$\Delta f/f_0$, showing that the trap fluctuations do not affect the magnitude of attractive forces. 
The data are compatible with a power-law exponent between $1$ and~$3/2$. 
Increasing the acoustic energy in the trap via the driving voltage of the transducer also proportionally increases the ratio~$\gamma/N_0$, confirming that the observed effective surface tension is a direct product of acoustic scattering forces (Fig.~\ref{fig:st}(a) inset). 

Unlike molecular liquids, our granular rafts can respond to tensile stresses by dilating slightly, and can thus respond elastically to an increase in rotational kinetic energy as long as~$\Sigma< \Sigma_\mathrm{max}$. 
In this regime, the change in raft perimeter is less than~$0.1$ times the raft radius~$R_\mathrm{eff}$, such that the surface tension contribution to this elastic expansion is negligible.
In Fig.~\ref{fig:drops}(d) this slight isotropic expansion takes place before the upturn in~$R^*$ and can be used to extract the effective {elastic modulus}. 
For our rotating rafts, which we treat as discs with thickness~$2a$ and circular face area~$A$, the fractional increase in volume~$V = 2aA$ from rest volume~$V_0=2aA_0$ in response to an applied rotational pressure is

\begin{align}
    \frac{V}{V_0}= 1+ \frac{1}{2K}\rho \omega^2 R_\mathrm{eff}^2 \, ,
    \label{eq:elasticmod}
\end{align}
where we have assumed that the granular material is linearly extensible with effective {elastic modulus}~$K$, {which we define in analogy to the bulk modulus of a three-dimensional material},  and particle material density~$\rho$ (see Appendix A for details, and Appendix C for examples of the data used to extract~$K$). 
Our measurements of~$K$, plotted in Fig.~\ref{fig:st}(b), show that the { effective} {elastic modulus} of the levitated granular rafts also scales with the number of constituent particles~$N_0$ (but not with the detuning parameter, within the scatter of the data). 
Increasing the acoustic energy density of the trap also increases the ratio~$K/N_0$ (Fig.~\ref{fig:st}(b) inset), confirming that the effective raft elasticity is directly controlled by the acoustic scattering forces between particles. 
For any driving amplitude, neither the  effective {elastic modulus} nor the effective surface tension appear to saturate, even up to rafts with 200 particles. If this size-dependence were to hold even in the thermodynamic limit, the effective {elastic modulus} and effective surface tension would not be well defined. 
However, we expect that the long-range cohesive forces holding the raft together would be screened at large length-scales, with a length-scale that depends on the geometry of the acoustic trap. For sufficiently large rafts, the surface tension and {elastic modulus} would then converge to intensive values that depend on the screening length.

{In the limit of very small droplets, the surface tension of small clusters of molecules and colloids at equilibrium also depends on the number of molecules in the droplet~\cite{nguyen2018measurement,lau2015water}. 
In these droplets, as the size decreases, the increasing curvature of the interface forces significant differences in the molecular structure, driving deviations in the surface tension from that of the elastic material. 
Such corrections become important when the size of this boundary layer becomes comparable to the droplet size. 
Previous work on small clusters in equilibrium has shown that these corrections come into play for three-dimensional clusters smaller than fifty particles, or clusters with a radius of approximately four particles. 
Such structural considerations thus seem unlikely to explain the size scaling observed here, where the effective surface tension does not appear to saturate up to rafts whose radii are greater than ten particles.}

Our results instead show that the acoustic binding energy itself scales with the raft size.
It is informative to compare this measured scaling of elastic constants with raft size to the results obtained by a pairwise acoustic scattering calculation. 
In particular, prior work suggested that in the Rayleigh limit ($a \ll \lambda$), the acoustic potential due to the presence of many particles can be calculated using a mean field approximation: the total acoustic potential {on a probe particle} is the pairwise (linear) sum of the potential due to each {source} particle~\cite{silva2014acoustic,zhang2016acoustically}. {In this approximation, source particles do not scatter sound previously scattered by other source particles, and are therefore considered as independent acoustic scatterers.}

However, the predicted {mean field} scaling of effective elastic constants with raft size (dashed lines in Fig.~\ref{fig:st}(a) and (b)) appears to have a power smaller than that observed in the experimental data (see Appendix A for details of the calculation). {For very small rafts, our data extrapolates to the mean field prediction. However, as the number of particles in a raft increases, the predicted scaling flattens out, while the experimental data do not.}
This discrepancy suggests that for large, close-packed rafts (more than {10} constituents), multibody forces ({from multiple scattering events}) contribute strongly to the total secondary acoustic potential, even if the individual constituents are well within the Rayleigh limit. 
{We note that the addition of an acoustic screening term would effectively reduce the number of source particles contributing to the total acoustic field at the probe particle, since screening introduces a length scale beyond which particle interactions are negligible. On the contrary, our results show that acoustic binding does not plateau as raft size is increased. Therefore screening effects, if present, occur at longer length scales than probed here. }

These {non-pairwise effects point to the fact that the rafts have entered the regime where source particles can no longer be treated as independent acoustic scatterers. In this regime, the non-additive forces could arise from} significant higher-order scattering between the close-packed particles, significant phase delays in the acoustic field between different parts of the raft, or alterations to the modal structure of the acoustic cavity due to the presence of the raft. 
Ultimately, our data highlight the current lack in understanding of secondary acoustic forces beyond lowest-order scattering expansions. 

As a counterpart to our observations on the overall shape-change of the rafts, we turn now to the size-dependence of micro-structural deformations in the rafts. 
Since the rafts allow for direct visual access to the configuration of individual constituent particles, we are able to track the micro-structural basis of their deformation throughout the course of their spin-up.
Our results, plotted in the upper half of Fig.~\ref{fig:st}(c), reveal that the largest rafts deform without changing their average connectivity: rafts composed of more than 100 particles rearrange their interior, without
changes in the average coordination number~$\langle z\rangle$, similar to sheared colloidal crystals~\cite{schall2004visualization,suenaga2007imaging}. 
However, as the rafts decrease in size, they deform by decreasing the average number of neighbors between particles in the elastic. 
This loss of stability in the raft interior results in changes in the mode of deformation past yielding. 
Plotting the fractional number of particles lost by the raft,~$N_\mathrm{loss}/N_0$, as it deforms (Fig.~\ref{fig:st}(d), example images in Fig.~\ref{fig:drops}(a) and insets of Fig.~\ref{fig:drops}(d)) reveals that small rafts tend to change shape by shedding particles rather than through rearrangements of their interior: small rafts appear brittle rather than ductile, as prefigured by the overall loss of connections in the interior of small rafts before their eventual failure.

The existence of such a cross-over from brittle to ductile behavior is in line with general considerations based on the relative {size-dependent} energetic costs of plastic deformation and fracture for small rafts. 
The cost of plastic deformation is, at a minimum, the energy required to create a dislocation pair in a previously crystalline domain, which scales as~$E\ln{R_\mathrm{eff}}$~\cite{chaikin1995principles}, where~$E$ is the Young's modulus of the material. 
Assuming that~$E$ and~$K$ scale similarly with~$N_0$ for our levitated granular rafts, the energy cost of a dislocation pair scales as~$N_0 \ln{N_0}$.
The actual cost may be higher: measurements of the flow stress for metallic nanopillars suggest that the energy for plastic deformation increases drastically for small crystals, as they may have an initially low dislocation density~\cite{greer2006nanoscale,jang2010transition}, or else rapidly exhaust their available dislocation sources~\cite{shan2008mechanical}.
On the other hand, the energetic cost of fracture is the energy required to create a new interface (whose {area} is on the order of the raft radius {times the particle diameter~$2a$}), which scales as~$\gamma N_0^{1/2}$ or, using the essentially linear dependence of~$\gamma$ on $N_0$, as $N_0^{3/2}$. 
For small~$N_0$ it is therefore favorable to fracture rather than to create dislocations. 
{The size-dependence of the raft elastic properties thus also results in differences in the micro-structural modes of rearrangement. }

\section{Conclusions}

We have used acoustic levitation to contactlessly assemble, drive, and measure the mechanics of active granular rafts. 
Here, these ``soft" granular rafts have attractive forces comparable to applied rotational tensions.
{Acoustic rotation thus offers the opportunity to tune through a wide range of driven behaviors in an inertial soft solid, from isotropic dilation, to extreme shape change and finally catastrophic failure.}

The observed size-dependence of the effective surface tension and {elastic modulus} of the rafts pose a particular challenge to the theoretical modeling of secondary acoustic forces. 
Such modeling~\cite{silva2014acoustic, sepehrirahnama2015numerical} currently relies on perturbative scattering expansions, which are appropriate in the limit of dilute Rayleigh scatterers, {where sound is scattered once between independent particles. These assumptions are valid in the regime} where individual particles are spaced far apart compared to the particle size, and are much smaller than the sound wavelength.  
However, these assumptions do not capture the dependence of the effective surface tension on particle number that we find for close-packed rafts, which can reach a sizeable fraction of the sound wavelength. 
Our results suggest the need for a systematic theoretical exploration of regimes in particle size and packing density in which acoustic interactions can no longer be treated as pairwise. 
This size-dependence also plays a role in governing both the elasticity and plasticity of these acoustic solids. 
In particular, the size range of our rafts spans the transition where small rafts deform by fracturing into pieces (similar to brittle failure), while larger rafts can respond to external stresses by plastically deforming their interior. {Our results demonstrate how acoustically levitated rafts can be used to investigate the mechanical properties of solids bound by non-pairwise interactions.
Consequences of non-pairwise forces for defect-mediated plasticity have been theorized in other systems~\cite{holian1991effects,baskes1999many,ziegenhain2009pair}, but are diﬃcult to observe experimentally.}

In addition, the size-dependence of the effective elasticity and cohesion of our levitated rafts is highly reminiscent of gravitationally-bound granular objects, where power-law gravitational forces lead to attractive forces that increase with the object size.
This resemblance to gravitational forces could open the door to a more detailed understanding of the dynamics of other rapidly rotating objects, such as rubble-pile asteroids. \edits{Such asteroids are generally understood to be granular aggregates bound by self-gravity~\cite{walsh2018rubble,hestroffer2019small,kollmer2021probing}. From these rubble piles, fission by rotation is thought to be a pathway to the creation of small binary asteroids~\cite{walsh2008rotational,jacobson2011dynamics}. The shape change of these bodies in response to rapid rotation is usually modelled by coarse-grained simulations of granular material held together by cohesive forces~\cite{sanchez2012simulation}. 
A key feature of these rubble-pile asteroids is an expansion of their equatorial cross-section as a consequence of increasing rotation rate, followed eventually by shape-change and break-up, similar to what we find in our rafts. 
Direct confirmation of these simulations is limited to observational studies, and more recently, the in-situ study of a few near-earth objects. Our granular rafts, whose cohesive forces scale similarly to gravitational binding, may thus serve as a coarse-grained tabletop model system for the dynamic evolution of the equatorial plane in granular astronomical objects that are rapidly rotating.}

\section{Acknowledgments}
We thank David Grier, Grayson Jackson, Tali Khain, Adam Kline, Vincenzo Vitelli, and Tom Witten for useful and inspiring discussions. This research was supported by the National Science Foundation through Grants No. DMR-1810390 and DMR-2104733.
This work utilized the shared experimental facilities at the University of Chicago MRSEC, which is funded by the National Science Foundation under award number DMR-2011854. 
This research utilized computational resources and services supported by the Research Computing Center at the University of Chicago. A.S. gratefully acknowledges the support of the Engineering and Physical Sciences Research Council (EPSRC) through New Investigator Award No. EP/T000961/1.
\section{Appendices}
\subsection*{Appendix A: Acoustic potential due to a disc composed of point particles}

\begin{figure}[h!]
    \centering
    \includegraphics[width = 0.5\columnwidth]{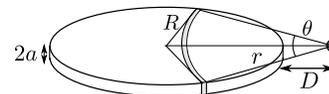}
    \caption{Schematic of the coordinate system for calculating the potential on a point particle due to a disc composed of point scatterers. The point particle, which has radius~$a$, is placed in plane with the disc (which has radius~$R$ and thickness~$2a$), and at a horizontal distance~$D$ from the edge of the disc. }
    \label{fig:SB_integrate}
\end{figure}

We consider the potential force on a point (test) particle due to a disc composed of point scatterers with radius~$a$ (such that the disc has height~$2a$), each of which has a pairwise interaction with the test particle. {For a disk composed of a monolayer packing of particles, the number density of point scatterers is~$\rho = 1/2\pi a^3$}. We compute the total potential~$U$ on the point particle as as the sum of the potentials~$u$ due to the disc constituents. This pairwise acoustic potential~$u$, plotted in Fig.~2 of the main text, is azimuthally symmetric, and depends on the radial distance between a pair of particles~$r$, as well as the polar angle between them,~$\psi$. 

For a test point placed at horizontal displacement~$D$ and zero vertical displacement from the edge of the disc (see Fig.~\ref{fig:SB_integrate} for a schematic), the set of points on the disc that are a distance~$r$ from the test point forms an arc, with arclength~$\theta$ and infinitesimal volume~$a r \, dr \,\theta$. Using the cosine rule, we have

\begin{align*}
    R^2 &= r^2 +(R+D)^2-2r(R+D)\cos{(\theta /2)}\\
    &\implies \theta = 2\arccos{\left(\frac{r^2+2RD+D^2}{2r(R+D)}\right)}.\
\end{align*}
Within this arc-volume, all constituents are at distance~$r$ from the test particle. The total acoustic potential due to the disc can thus be derived by integrating the contributions of each arc-length volume over the area of the disc. This can be expressed as a one-dimensional integral over~$r$: 

\begin{align}
\begin{split}
    U(D)= 4a\rho\int_D^{2R+D} dr\: u(r) r \times \\
    \arccos{\left( \frac{r^2+2RD+D^2}{2r(R+D)}\right)} 
    \end{split}
\end{align}

We carry out this integration numerically for different values of~$R$,~{with~$D=2a$}, to produce the dashed line in Fig.~4(a) of the main text. The mean-field prediction for the acoustic {elastic modulus} (dashed line in Fig.~4(b)) is the second derivative of this expression with respect to the coordinate~$D$, {evaluated at~$D=2a$}, which we also carry out numerically. 

\subsection*{Appendix B: Observation number}

Table 1 lists the number of observations of spinning dimers that were combined for the data in Fig.~2(d) and~(e) of the main text. 

\begin{table}[h!]
\centering 
\begin{tabular}{c c}
\hline
\hline                  
$\Delta f/f_0\times 10^{-3}$ & number of observations \\[0.5ex]
\hline
0.46 & 17  \\
0.89 & 32  \\
0.96 & 26  \\
1.77 & 20  \\
2.21 & 20  \\
2.87 & 35  \\
3.53 & 30  \\[1ex]
\hline
\end{tabular}
\caption{Number of dimer pairs observed for Fig. 2(d) and~(e). Each dimer was observed for 23 seconds.}
\end{table}

\subsection*{Appendix C: Measurement of effective {elastic modulus}}

\begin{figure}[h!]
    \centering
    \includegraphics[width = 1\columnwidth]{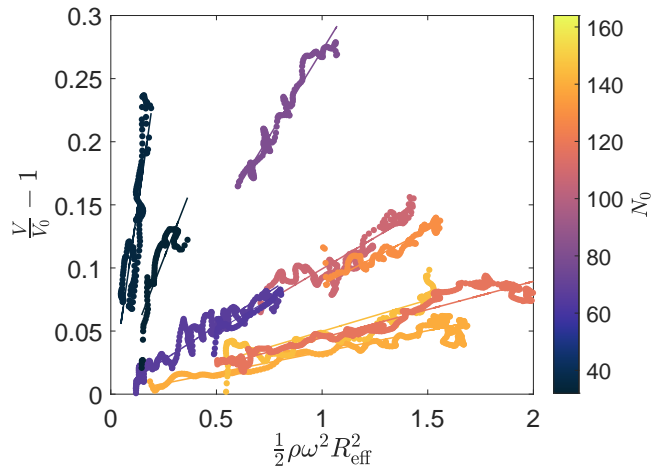}
    \caption{Example raw data for the calculation of~$K$ (dots). The fractional change in volume of the raft~$V/V_0-1$ is plotted as a function of the rotational pressure,~$\frac{1}{2}\rho \omega^2 R_\mathrm{eff}^2$. Color indicates~$N_0$. The best fit line is indicated as a solid line through each data set. Data has been smoothed with a moving average filter, with width 10 data points (3ms).}
    \label{fig:compress_raw}
\end{figure}

In order to make use of Eq.~\ref{eq:elasticmod} and extract the effective {elastic modulus}~$K$ of the droplets, we plot the fractional change in raft volume~$V/V_0-1$ {(where $V = aA$, the product of the circular face area of the raft and its thickness)} as a function of the rotational pressure~$\frac{1}{2}\rho \omega^2 R_\mathrm{eff}^2$ (example data shown in Fig.~\ref{fig:compress_raw}).  Here, we determine~$V_0$, or equivalently, the initial area of the rafts (when $\omega=0$) {times their thickness}, by fitting the data to lines, and extracting the intercept. The data is then fitted to a line, whose slope then gives~$1/K$. Error bars in Fig.~\ref{fig:st}(b) reflect the least-squares error in the fit coefficients. 

\subsection*{Appendix D: Plasticity during raft deformation}

\begin{figure}
    \centering
    \includegraphics[width = 1\columnwidth]{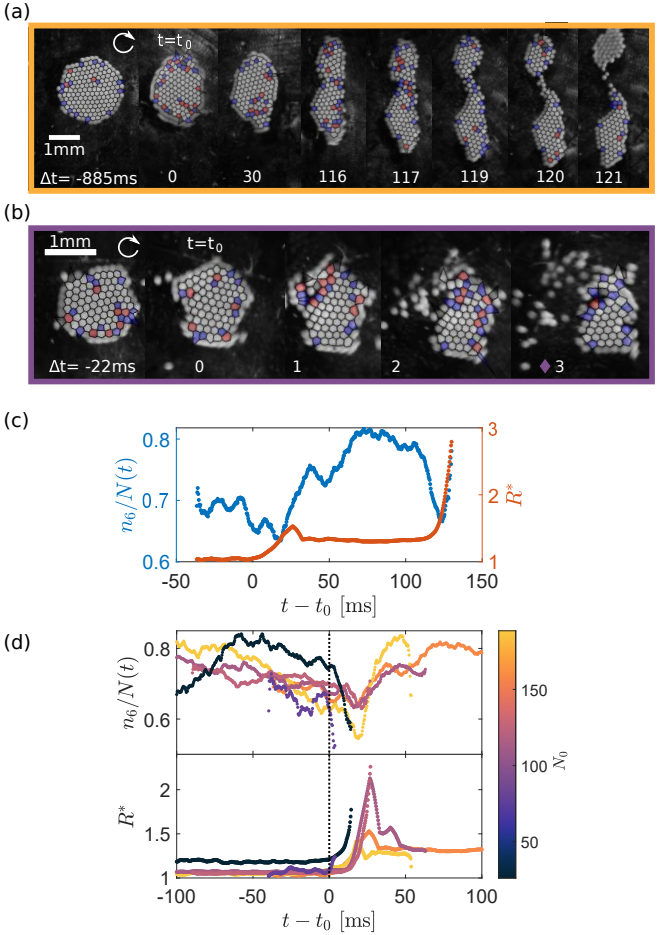}
    \caption{Voronoi construction reveals the onset of plasticity during raft deformation. (a,b) Images in Fig 3(a) of the main text, overlaid with the corresponding Voronoi diagram of the raft interior. Cells in the Voronoi diagram are overlaid with the corresponding number of sides of the polygon: particles with five neighbours are shaded blue, particles with six neighbours white, and particles with seven neighbours red. (c) (left axis) Plot of the number of particles in the raft interior with six neighbours~$n_6$, divided by the total number of particles in the raft~$N(t)$, as a function of time (both~$n_6$ and~$N$ are tracked at every frame of the drop evolution), for the raft pictured in (a). (right axis) Raft shape parameter~$R^*$ as a function of time. (d) (upper) $n_6/N(t)$ as a function of time, for several different initial raft sizes~$N_0$. (lower) $R^*$ as a function of time.}
    \label{fig:voro}
\end{figure}

In order to assess the relative roles of plasticity and elasticity during the raft deformation past~$\Sigma_\mathrm{max}$, we measure the crystallinity of the raft interior (via the particle coordination number, measured using a Voronoi diagram) throughout the course of its rotational break-up. 
Example snapshots of the Voronoi diagrams are shown in Fig.~\ref{fig:voro}(a) and (b), at different points throughout the raft deformation. 
The Voronoi statistics for the raft in Fig.~\ref{fig:voro}(a) are summarized in Fig.~\ref{fig:voro}(c), which plots the evolution of the number of particles with six neighbours in the raft interior~$n_6$, as a fraction of the total number of particles in the raft~$N(t)$, together with the shape parameter~$R^*$ as a function of time. At first, the shape of the raft does not change (constant~$R^*$). At the same time, in response to the growing spinning speed,~$n_6$ slowly decreases. As the rotational speed of the raft increases, a small number of defects are accumulated in the interior. During this phase of raft spin-up, the stresses are predominantly dilational, which do not motivate the glide of dislocations. 

The shape of the raft then changes modestly, resulting in an also modest drop in~$n_6$ around 30ms. After this point, the raft retains its new elongated shape for a considerable period of time. During this period of time, $n_6$ grows, and in fact exceeds the value of~$n_6$ observed for the initially circular raft: the crystal structure of the raft is now able to relax and remove defects, in a form of rotational annealing. The elongation of the raft into an ellipse breaks the azimuthal symmetry of the rotational stress, introducing shear fields to the raft interior which sweep dislocations to the raft boundaries. Finally, the rotational energy again exceeds the binding energy of the drop, which then elongates to~$R^*=3$, and pinches off into two drops. This extreme elongation rapidly generates defects and drastically lowers~$n_6$. Again, after the pinch-off,~$n_6$ rises, indicating that the remaining two sections of the raft have eliminated their defects through the creation of additional surface, and now have well-ordered interiors.  

Further examination of~$n_6/N(t)$ and~$R^*$ as a function of time, for several different raft sizes, (Fig.~\ref{fig:voro}(d)) confirms these trends for large rafts: raft shape changes are accompanied by a temporary decrease in~$n_6$, which then recovers after a short period in which the raft crystal structure relaxes.  In contrast, since the smallest rafts tend to change shape by shedding large fractions of their constituent particles,~$n_6/N(t)$ drops sharply at~$t_0$, and does not return to a close-packed state after the droplet shape change. In all cases, we observe a gradual decrease in~$n_6/N(t)$ during the spin-up process, indicating that the dilation of the raft before global-shape change generates defects in the raft interior. This gradual decrease is marked by fluctuations that correspond to the appearance and disappearance of several defect pairs.  

{

\subsection*{Appendix E: Nonaffine particle motion}
\begin{figure}
    \centering
    \includegraphics[width = 0.67\columnwidth]{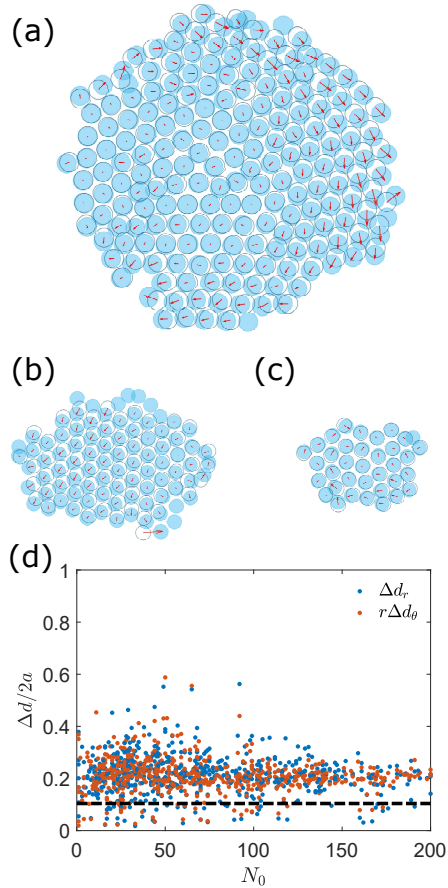}
    \caption{{Nonaffine motion of particles before breakup. (a--c) Comparison of rotation-corrected particle positions at the start of spin-up (dark blue circles), and just before shape change (i.e. one frame before~$E_\mathrm{rot}=E_\mathrm{rot,max}$, plotted as light blue filled circles), for three rafts of varying size. Red arrows show the local displacement field from initial to final particle position. (d) Plot of average displacement of particles (averaged over all particles in the raft) during spin-up~$\Delta d$ as a function of~$N_0$, normalized by the particle diameter~$2a$. Displacement has been resolved into the radial ($\Delta d_r$, blue data points), and azimuthal ($r\Delta d_\theta$, orange data points) directions. Black dashed line indicates the noise floor for particle tracking.}}
    \label{fig:rigidbody}
\end{figure}

One of the assumptions underlying the derivation of Eq.~\ref{eq:gamma} is that the raft rotates as a rigid body up until the point where it changes shape, i.e. that there is negligible non-affine motion of the particles in the raft for~$E_\mathrm{rot}<E_\mathrm{rot,max}$. 
In order to assess the validity of this assumption, we measure the position of particles in the raft at the start and end of the spin-up process (the end of the spin-up process defined as one frame, or $0.33$ms, before the raft reaches its maximum rate of rotation). 
Once we correct for the rigid-body translation and rotation of the cluster, we then construct the displacement vector between all matching particles in the raft.

Example particle positions and displacements are shown in Fig.~\ref{fig:rigidbody}(a--c): the vast majority of particle displacements have magnitude smaller than a particle radius.
Quantitatively, we resolve the displacements into the radial and azimuthal directions, and plot the mean directional magnitudes (averaged over all particles in a raft) as a function of~$N_0$ (Fig.~\ref{fig:rigidbody}(d)). 
We find that the average displacements in the radial and tangential directions are of similar magnitude, roughly half a particle radius. This is the case even in the larger rafts, where there can be some long-wavelength collective motion as the raft dilates (see, e.g., Fig.~\ref{fig:rigidbody}(a)). However, the magnitude of this motion stays well below a lattice spacing (the membership of particle nearest neighbor shells is not disturbed). Thus, the rafts remain solid-like, rigid, and do not exhibit evidence of internal shear before the onset of deformation.

\subsection*{Appendix F: Estimate of Stokes drag on raft perimeter}

\begin{figure}
    \centering
    \includegraphics[width = 0.9\columnwidth]{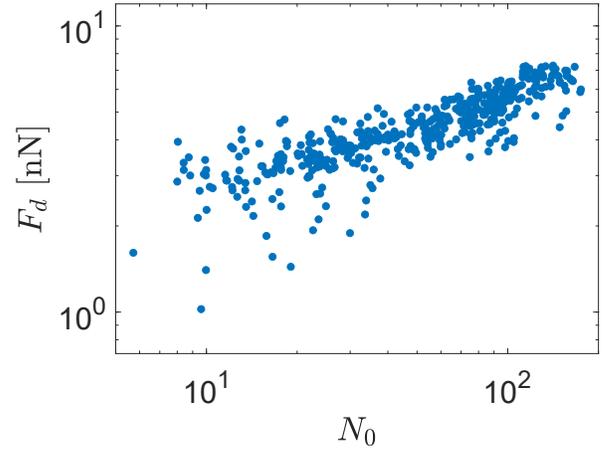}
    \caption{{Drag force~$F_d$ on a particle on the perimeter of a raft, estimated using a linear drag model, as a function of the total number of particles in the raft~$N_0$.}}
    \label{fig:stokesN0}
\end{figure}

In order to assess the effect of air drag and compare it to the acoustic binding force between particles, we calculate the viscous force on a particle at the edge of the raft, using a linear drag model: 
\begin{align*}
    F_d=6\pi\eta a v
\end{align*}
where~$\eta$ is the viscosity of air,~$a$ is the radius of an individual particle, and~$v=R_\mathrm{eff}\omega$ is the linear velocity of a particle on the perimeter of a raft with radius~$R_\mathrm{eff}$, rotating with angular velocity~$\omega$. 
This model for~$F_d$ (measured at the moment of shape change, where~$\omega$ is largest, yields measurements that are of order 1-10 nN (Fig.~\ref{fig:stokesN0}).

For comparison, we need to estimate the magnitude of the secondary acoustic forces between a pair of particles that compose the raft. One estimate is provided by the centripetal force required to keep the particle attached to the edge of the raft,~$F_c = m\omega^2 R_\mathrm{eff}$. 
Since the particles do not detach from the raft, such an estimate serves as a lower bound on the secondary acoustic force acting on a single particle. 
The ratio of this centripetal force to the drag force is~$F_c/F_d = \frac{2\rho \omega a^2}{9\eta} \approx 50$. 
Alternatively, Fig. 2(b) shows that the radial restoring force between a pair of particles is approximately 0.1, in units of $E_0 V_p/a$, where $E_0$ is the energy density of the cavity, $V_p$ is the particle volume, and $a$ is the particle radius. To convert the values in Fig.~2(b) to a force in Newtons, we infer the energy density from the measurement of surface tension. Fig.~4(a) compares the surface tension, measured from the experiments, to an analytical calculation obtained by integrating the acoustic potential $U$ over a disk (black dotted line). Using $\gamma \sim U/a^2$, the fit between this analytical calculation and the experimental data has a single fitting parameter, corresponding to~$E_0$. Substituting this value of~$E_0$, we find that the secondary acoustic force for a particle on the edge of the smallest rafts (10-30 particles) is of order 1$\mu$N, or two to three orders of magnitude greater than the estimated drag force. Fitting the analytical calculation to the surface tension of the largest rafts increases the estimated secondary acoustic forces to roughly 100$\mu$N. Again, this estimates provides a lower bound on the acoustic binding between particles in the raft. The raft shape changes and thus the data on the effective surface tension and {elastic modulus} are thus unlikely to be affected by air drag.

\subsection*{Appendix G: Rotationally activated out-of-plane bending}

\begin{figure}
\centering
\includegraphics[width = 1\columnwidth]{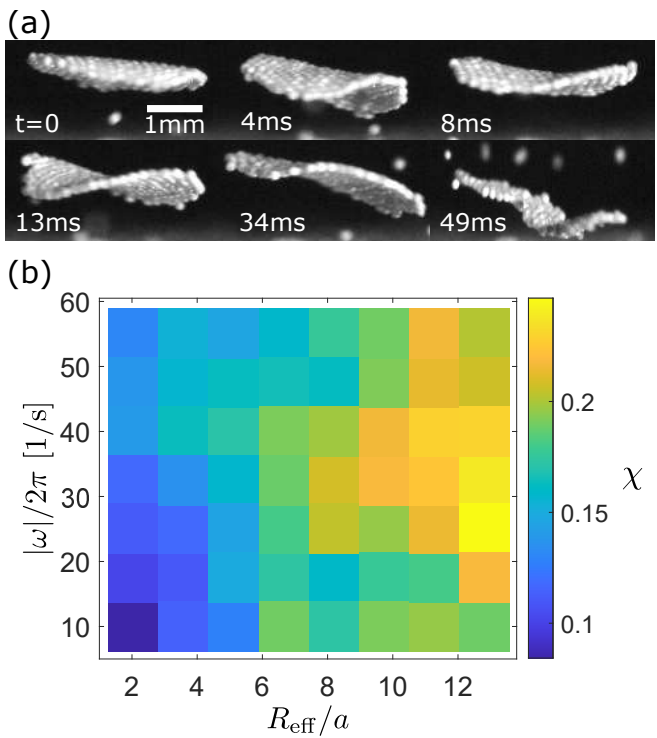}
\caption{Rotation activates out-of-plane bending in levitated granular rafts. 
 (a) Sequence of side-view images showing the rotation of a raft, which eventually elongates into an ellipse (t=34ms).
During the process of spinning up, the raft deforms significantly out of plane. (b) Raft concavity~$\chi$ (value shown by colorbar, see text for definition), as a function of the absolute rotational frequency~$|\omega|$ and raft radius~$R_\mathrm{eff}/a$. The data represents observations of 60 (independently self-assembled) rafts. The 95\% confidence intervals for~$\chi$ are approximately constant as a function of~$|\omega|$ and~$R_\mathrm{eff}/a$, and do not exceed 0.02 for any measurement. }
\label{fig:boutplane}
\end{figure}

While the effective in-plane surface tension and effective {elastic modulus} of the rafts are controlled by the in-plane component of the secondary acoustic forces (Fig.~\ref{fig:setup}(a)), the rafts have other elastic constants for out-of-plane deformations, which are controlled by the out-of-plane components of the secondary acoustic force (Fig.~\ref{fig:setup}(b)). 
As a result of this out-of-plane bending modulus and their membrane-like, effectively two-dimensional nature, levitated granular rafts can strongly deform out of plane (Fig.~\ref{fig:boutplane}(a), see Supplementary Movie 4 for dynamics). 
The smallest rafts remain planar as they rotate, exhibiting rocking motions that tilt the raft in and out of the levitation plane. 
However, for the largest rafts, increasing the rotation rate of the rafts gives rise to saddle-like bending modes, which can start waves that travel azimuthally around the raft while it rotates. 

{In the following, we discuss these transverse bending modes and waves at raft rotation speeds in the regime where the rafts remain circular (before they deform and break up).} To characterize these waves, we make use of the fact that the bending modes appear as saddle-like, nonconvex structures when viewed from the side. We thus use the concavity of the (thresholded) black and white side-view images as a proxy for the out-of-plane bending of the raft. 
Given a raft shape with projected area~$A_I$, whose corresponding convex hull has area~$A_v$, we define the concavity~$\chi$ as~$\chi=1-A_I/A_v$. 
For planar objects viewed from any angle, the convex hull of the image is almost completely filled by the original image, and so~$\chi$ is close to zero. 

Measurements of~$\chi$ as a function of the absolute rotation rate~$|\omega|$ and normalized raft radius~$R_\mathrm{eff}/a$ reveal that rotation activates out-of-plane bending nonmonotonically (Fig.~\ref{fig:boutplane}(b)). The smallest rafts remain relatively flat as they rotate faster, with a weak increase of~$\chi$.
For larger~$R_\mathrm{eff}/a$, more out-of-plane degrees of freedom are available, and the overall concavity at first increases with~$|\omega|$, up to~$|\omega|/2\pi \simeq$ 30Hz, but then decreases for faster spinning. 

{Similar out-of-plane bending modes occur in thin, rapidly spinning elastic disks and membranes, where they originate from nonlinear interactions between the shape of an elastic object, and the shape-dependent aerodynamic drag on it. These nonlinear interactions result not only in a drag force, but also an additional lift force, whose magnitude is proportional to both the angular speed and the out-of-plane displacement of the raft~\cite{yasuda1992self}. Even a very small lift force (compared to the forces holding the disk together) can then excite and amplify vibrational modes in the form of traveling waves around the disk edge~\cite{nowinski1964nonlinear,mote1993aerodynamically}. }

 Since the measured effective {elastic modulus} of our membrane-like rafts is an order of magnitude smaller than in most previously observed spinning disks~\cite{mote1993aerodynamically, renshaw1994aerodynamically, kang2006vibrations}, increasing~$\omega$ can tune through a wider range of {behaviors}.  In particular, {for the larger rafts we can reach a regime where the inertial forces during rotation increase the effective raft tension, to the point that this reduces the amplitude of transverse undulation}. 
This rotationally induced reduction in out-of-plane motion has been observed previously only in the limit of membranes that are extremely thin relative to their radius, such that the bending stiffness is negligibly small~\cite{okuizumi2007equilibrium,guven2013whirling,delapierre2018wrinkling}.

}

\subsection*{Appendix H: Description of Supplementary Movies}
\textbf{Supplementary Movie 1.} Movie showing the underdamped self-asssembly of polyethylene particles from a side-view. Particles are initially scattered on the reflector surface. When the acoustic field is turned on, the particles levitate, oscillating about the nodal plane. Driven by secondary scattering forces, particles at first form small clusters, which then coalesce to form a monolayer of particles. See Fig. 1(a) of the main text for corresponding still images. 

\textbf{Supplementary Movie 2.} Movie (taken from below) showing two rafts approaching each other and merging into a larger raft. Variations in brightness correspond to local curling of the droplet out of plane. See Fig 1(c) of the main text for corresponding still images. 

\textbf{Supplementary Movie 3.} Movies (taken from below) showing the deformation of two initially circular rafts that rotate in the clockwise direction. As the droplet gains angular momentum, it elongates into an ellipse, then splits into several smaller droplets. The pieces then collide and merge, before spinning up and splitting again. See Fig 3(a) and (b) of the main text for corresponding still images. 

\textbf{Supplementary Movie 4.} Short video to illustrate Fig. 10, beginning with a movie (taken from the side) showing the rotation of a raft, which eventually elongates into an ellipse. During the process of spinning up, the raft deforms significantly out of plane. See Fig 10(a) for corresponding still images. The video goes on to illustrate Fig 10(b) with movies of the rotation of rafts at different areas in the color plot. 


\begin{thebibliography}{73}%
\makeatletter
\providecommand \@ifxundefined [1]{%
 \@ifx{#1\undefined}
}%
\providecommand \@ifnum [1]{%
 \ifnum #1\expandafter \@firstoftwo
 \else \expandafter \@secondoftwo
 \fi
}%
\providecommand \@ifx [1]{%
 \ifx #1\expandafter \@firstoftwo
 \else \expandafter \@secondoftwo
 \fi
}%
\providecommand \natexlab [1]{#1}%
\providecommand \enquote  [1]{``#1''}%
\providecommand \bibnamefont  [1]{#1}%
\providecommand \bibfnamefont [1]{#1}%
\providecommand \citenamefont [1]{#1}%
\providecommand \href@noop [0]{\@secondoftwo}%
\providecommand \href [0]{\begingroup \@sanitize@url \@href}%
\providecommand \@href[1]{\@@startlink{#1}\@@href}%
\providecommand \@@href[1]{\endgroup#1\@@endlink}%
\providecommand \@sanitize@url [0]{\catcode `\\12\catcode `\$12\catcode
  `\&12\catcode `\#12\catcode `\^12\catcode `\_12\catcode `\%12\relax}%
\providecommand \@@startlink[1]{}%
\providecommand \@@endlink[0]{}%
\providecommand \url  [0]{\begingroup\@sanitize@url \@url }%
\providecommand \@url [1]{\endgroup\@href {#1}{\urlprefix }}%
\providecommand \urlprefix  [0]{URL }%
\providecommand \Eprint [0]{\href }%
\providecommand \doibase [0]{https://doi.org/}%
\providecommand \selectlanguage [0]{\@gobble}%
\providecommand \bibinfo  [0]{\@secondoftwo}%
\providecommand \bibfield  [0]{\@secondoftwo}%
\providecommand \translation [1]{[#1]}%
\providecommand \BibitemOpen [0]{}%
\providecommand \bibitemStop [0]{}%
\providecommand \bibitemNoStop [0]{.\EOS\space}%
\providecommand \EOS [0]{\spacefactor3000\relax}%
\providecommand \BibitemShut  [1]{\csname bibitem#1\endcsname}%
\let\auto@bib@innerbib\@empty
\bibitem [{\citenamefont {Smarr}(1973)}]{smarr1973mass}%
  \BibitemOpen
  \bibfield  {author} {\bibinfo {author} {\bibfnamefont {L.}~\bibnamefont
  {Smarr}},\ }\bibfield  {title} {\bibinfo {title} {Mass formula for {Kerr}
  black holes},\ }\href@noop {} {\bibfield  {journal} {\bibinfo  {journal}
  {Physical Review Letters}\ }\textbf {\bibinfo {volume} {30}},\ \bibinfo
  {pages} {71} (\bibinfo {year} {1973})}\BibitemShut {NoStop}%
\bibitem [{\citenamefont {Genzel}\ \emph {et~al.}(2003)\citenamefont {Genzel},
  \citenamefont {Sch{\"o}del}, \citenamefont {Ott}, \citenamefont {Eckart},
  \citenamefont {Alexander}, \citenamefont {Lacombe}, \citenamefont {Rouan},\
  and\ \citenamefont {Aschenbach}}]{genzel2003near}%
  \BibitemOpen
  \bibfield  {author} {\bibinfo {author} {\bibfnamefont {R.}~\bibnamefont
  {Genzel}}, \bibinfo {author} {\bibfnamefont {R.}~\bibnamefont {Sch{\"o}del}},
  \bibinfo {author} {\bibfnamefont {T.}~\bibnamefont {Ott}}, \bibinfo {author}
  {\bibfnamefont {A.}~\bibnamefont {Eckart}}, \bibinfo {author} {\bibfnamefont
  {T.}~\bibnamefont {Alexander}}, \bibinfo {author} {\bibfnamefont
  {F.}~\bibnamefont {Lacombe}}, \bibinfo {author} {\bibfnamefont
  {D.}~\bibnamefont {Rouan}},\ and\ \bibinfo {author} {\bibfnamefont
  {B.}~\bibnamefont {Aschenbach}},\ }\bibfield  {title} {\bibinfo {title}
  {Near-infrared flares from accreting gas around the supermassive black hole
  at the galactic centre},\ }\href@noop {} {\bibfield  {journal} {\bibinfo
  {journal} {Nature}\ }\textbf {\bibinfo {volume} {425}},\ \bibinfo {pages}
  {934} (\bibinfo {year} {2003})}\BibitemShut {NoStop}%
\bibitem [{\citenamefont {McKinney}\ and\ \citenamefont
  {Gammie}(2004)}]{mckinney2004measurement}%
  \BibitemOpen
  \bibfield  {author} {\bibinfo {author} {\bibfnamefont {J.~C.}\ \bibnamefont
  {McKinney}}\ and\ \bibinfo {author} {\bibfnamefont {C.~F.}\ \bibnamefont
  {Gammie}},\ }\bibfield  {title} {\bibinfo {title} {A measurement of the
  electromagnetic luminosity of a {Kerr} black hole},\ }\href@noop {}
  {\bibfield  {journal} {\bibinfo  {journal} {The astrophysical journal}\
  }\textbf {\bibinfo {volume} {611}},\ \bibinfo {pages} {977} (\bibinfo {year}
  {2004})}\BibitemShut {NoStop}%
\bibitem [{\citenamefont {Walsh}\ \emph {et~al.}(2008)\citenamefont {Walsh},
  \citenamefont {Richardson},\ and\ \citenamefont
  {Michel}}]{walsh2008rotational}%
  \BibitemOpen
  \bibfield  {author} {\bibinfo {author} {\bibfnamefont {K.~J.}\ \bibnamefont
  {Walsh}}, \bibinfo {author} {\bibfnamefont {D.~C.}\ \bibnamefont
  {Richardson}},\ and\ \bibinfo {author} {\bibfnamefont {P.}~\bibnamefont
  {Michel}},\ }\bibfield  {title} {\bibinfo {title} {Rotational breakup as the
  origin of small binary asteroids},\ }\href@noop {} {\bibfield  {journal}
  {\bibinfo  {journal} {Nature}\ }\textbf {\bibinfo {volume} {454}},\ \bibinfo
  {pages} {188} (\bibinfo {year} {2008})}\BibitemShut {NoStop}%
\bibitem [{\citenamefont {Rozitis}\ \emph {et~al.}(2014)\citenamefont
  {Rozitis}, \citenamefont {MacLennan},\ and\ \citenamefont
  {Emery}}]{rozitis2014cohesive}%
  \BibitemOpen
  \bibfield  {author} {\bibinfo {author} {\bibfnamefont {B.}~\bibnamefont
  {Rozitis}}, \bibinfo {author} {\bibfnamefont {E.}~\bibnamefont {MacLennan}},\
  and\ \bibinfo {author} {\bibfnamefont {J.~P.}\ \bibnamefont {Emery}},\
  }\bibfield  {title} {\bibinfo {title} {Cohesive forces prevent the rotational
  breakup of rubble-pile asteroid (29075) 1950 {DA}},\ }\href@noop {}
  {\bibfield  {journal} {\bibinfo  {journal} {Nature}\ }\textbf {\bibinfo
  {volume} {512}},\ \bibinfo {pages} {174} (\bibinfo {year}
  {2014})}\BibitemShut {NoStop}%
\bibitem [{\citenamefont {Barnouin}\ \emph {et~al.}(2019)\citenamefont
  {Barnouin}, \citenamefont {Daly}, \citenamefont {Palmer}, \citenamefont
  {Gaskell}, \citenamefont {Weirich}, \citenamefont {Johnson}, \citenamefont
  {Al~Asad}, \citenamefont {Roberts}, \citenamefont {Perry}, \citenamefont
  {Susorney} \emph {et~al.}}]{barnouin2019shape}%
  \BibitemOpen
  \bibfield  {author} {\bibinfo {author} {\bibfnamefont {O.}~\bibnamefont
  {Barnouin}}, \bibinfo {author} {\bibfnamefont {M.}~\bibnamefont {Daly}},
  \bibinfo {author} {\bibfnamefont {E.}~\bibnamefont {Palmer}}, \bibinfo
  {author} {\bibfnamefont {R.}~\bibnamefont {Gaskell}}, \bibinfo {author}
  {\bibfnamefont {J.}~\bibnamefont {Weirich}}, \bibinfo {author} {\bibfnamefont
  {C.}~\bibnamefont {Johnson}}, \bibinfo {author} {\bibfnamefont
  {M.}~\bibnamefont {Al~Asad}}, \bibinfo {author} {\bibfnamefont
  {J.}~\bibnamefont {Roberts}}, \bibinfo {author} {\bibfnamefont
  {M.}~\bibnamefont {Perry}}, \bibinfo {author} {\bibfnamefont
  {H.}~\bibnamefont {Susorney}}, \emph {et~al.},\ }\bibfield  {title} {\bibinfo
  {title} {Shape of (101955) {Bennu} indicative of a rubble pile with internal
  stiffness},\ }\href@noop {} {\bibfield  {journal} {\bibinfo  {journal}
  {Nature Geoscience}\ }\textbf {\bibinfo {volume} {12}},\ \bibinfo {pages}
  {247} (\bibinfo {year} {2019})}\BibitemShut {NoStop}%
\bibitem [{\citenamefont {Arita}\ \emph {et~al.}(2013)\citenamefont {Arita},
  \citenamefont {Mazilu},\ and\ \citenamefont {Dholakia}}]{arita2013laser}%
  \BibitemOpen
  \bibfield  {author} {\bibinfo {author} {\bibfnamefont {Y.}~\bibnamefont
  {Arita}}, \bibinfo {author} {\bibfnamefont {M.}~\bibnamefont {Mazilu}},\ and\
  \bibinfo {author} {\bibfnamefont {K.}~\bibnamefont {Dholakia}},\ }\bibfield
  {title} {\bibinfo {title} {Laser-induced rotation and cooling of a trapped
  microgyroscope in vacuum},\ }\href@noop {} {\bibfield  {journal} {\bibinfo
  {journal} {Nature Communications}\ }\textbf {\bibinfo {volume} {4}},\
  \bibinfo {pages} {1} (\bibinfo {year} {2013})}\BibitemShut {NoStop}%
\bibitem [{\citenamefont {Cohen}\ \emph {et~al.}(1974)\citenamefont {Cohen},
  \citenamefont {Plasil},\ and\ \citenamefont
  {Swiatecki}}]{cohen1974equilibrium}%
  \BibitemOpen
  \bibfield  {author} {\bibinfo {author} {\bibfnamefont {S.}~\bibnamefont
  {Cohen}}, \bibinfo {author} {\bibfnamefont {F.}~\bibnamefont {Plasil}},\ and\
  \bibinfo {author} {\bibfnamefont {W.}~\bibnamefont {Swiatecki}},\ }\bibfield
  {title} {\bibinfo {title} {Equilibrium configurations of rotating charged or
  gravitating liquid masses with surface tension. {II}},\ }\href@noop {}
  {\bibfield  {journal} {\bibinfo  {journal} {Annals of Physics}\ }\textbf
  {\bibinfo {volume} {82}},\ \bibinfo {pages} {557} (\bibinfo {year}
  {1974})}\BibitemShut {NoStop}%
\bibitem [{\citenamefont {Pomorski}\ and\ \citenamefont
  {Dudek}(2003)}]{pomorski2003nuclear}%
  \BibitemOpen
  \bibfield  {author} {\bibinfo {author} {\bibfnamefont {K.}~\bibnamefont
  {Pomorski}}\ and\ \bibinfo {author} {\bibfnamefont {J.}~\bibnamefont
  {Dudek}},\ }\bibfield  {title} {\bibinfo {title} {Nuclear liquid-drop model
  and surface-curvature effects},\ }\href@noop {} {\bibfield  {journal}
  {\bibinfo  {journal} {Physical Review C}\ }\textbf {\bibinfo {volume} {67}},\
  \bibinfo {pages} {044316} (\bibinfo {year} {2003})}\BibitemShut {NoStop}%
\bibitem [{\citenamefont {Schunck}\ \emph {et~al.}(2007)\citenamefont
  {Schunck}, \citenamefont {Dudek},\ and\ \citenamefont
  {Herskind}}]{schunck2007nuclear}%
  \BibitemOpen
  \bibfield  {author} {\bibinfo {author} {\bibfnamefont {N.}~\bibnamefont
  {Schunck}}, \bibinfo {author} {\bibfnamefont {J.}~\bibnamefont {Dudek}},\
  and\ \bibinfo {author} {\bibfnamefont {B.}~\bibnamefont {Herskind}},\
  }\bibfield  {title} {\bibinfo {title} {Nuclear hyperdeformation and the
  {Jacobi} shape transition},\ }\href@noop {} {\bibfield  {journal} {\bibinfo
  {journal} {Physical Review C}\ }\textbf {\bibinfo {volume} {75}},\ \bibinfo
  {pages} {054304} (\bibinfo {year} {2007})}\BibitemShut {NoStop}%
\bibitem [{\citenamefont {Arabgol}\ and\ \citenamefont
  {Sleator}(2019)}]{arabgol2019observation}%
  \BibitemOpen
  \bibfield  {author} {\bibinfo {author} {\bibfnamefont {M.}~\bibnamefont
  {Arabgol}}\ and\ \bibinfo {author} {\bibfnamefont {T.}~\bibnamefont
  {Sleator}},\ }\bibfield  {title} {\bibinfo {title} {Observation of the
  nuclear {Barnett} effect},\ }\href@noop {} {\bibfield  {journal} {\bibinfo
  {journal} {Physical Review Letters}\ }\textbf {\bibinfo {volume} {122}},\
  \bibinfo {pages} {177202} (\bibinfo {year} {2019})}\BibitemShut {NoStop}%
\bibitem [{\citenamefont {Hill}\ and\ \citenamefont
  {Eaves}(2008)}]{hill2008nonaxisymmetric}%
  \BibitemOpen
  \bibfield  {author} {\bibinfo {author} {\bibfnamefont {R.}~\bibnamefont
  {Hill}}\ and\ \bibinfo {author} {\bibfnamefont {L.}~\bibnamefont {Eaves}},\
  }\bibfield  {title} {\bibinfo {title} {Nonaxisymmetric shapes of a
  magnetically levitated and spinning water droplet},\ }\href@noop {}
  {\bibfield  {journal} {\bibinfo  {journal} {Physical Review Letters}\
  }\textbf {\bibinfo {volume} {101}},\ \bibinfo {pages} {234501} (\bibinfo
  {year} {2008})}\BibitemShut {NoStop}%
\bibitem [{\citenamefont {Chandrasekhar}(1965)}]{chandrasekhar1965stability}%
  \BibitemOpen
  \bibfield  {author} {\bibinfo {author} {\bibfnamefont {S.}~\bibnamefont
  {Chandrasekhar}},\ }\bibfield  {title} {\bibinfo {title} {The stability of a
  rotating liquid drop},\ }\href@noop {} {\bibfield  {journal} {\bibinfo
  {journal} {Proceedings of the Royal Society of London. Series A. Mathematical
  and Physical Sciences}\ }\textbf {\bibinfo {volume} {286}},\ \bibinfo {pages}
  {1} (\bibinfo {year} {1965})}\BibitemShut {NoStop}%
\bibitem [{\citenamefont {Brown}\ and\ \citenamefont
  {Scriven}(1980)}]{brown1980shape}%
  \BibitemOpen
  \bibfield  {author} {\bibinfo {author} {\bibfnamefont {R.}~\bibnamefont
  {Brown}}\ and\ \bibinfo {author} {\bibfnamefont {L.}~\bibnamefont
  {Scriven}},\ }\bibfield  {title} {\bibinfo {title} {The shape and stability
  of rotating liquid drops},\ }\href@noop {} {\bibfield  {journal} {\bibinfo
  {journal} {Proceedings of the Royal Society of London. A. Mathematical and
  Physical Sciences}\ }\textbf {\bibinfo {volume} {371}},\ \bibinfo {pages}
  {331} (\bibinfo {year} {1980})}\BibitemShut {NoStop}%
\bibitem [{\citenamefont {Chen}\ \emph {et~al.}(2006)\citenamefont {Chen},
  \citenamefont {Shi}, \citenamefont {Zhang}, \citenamefont {Zhu},\ and\
  \citenamefont {Yan}}]{chen2006size}%
  \BibitemOpen
  \bibfield  {author} {\bibinfo {author} {\bibfnamefont {C.}~\bibnamefont
  {Chen}}, \bibinfo {author} {\bibfnamefont {Y.}~\bibnamefont {Shi}}, \bibinfo
  {author} {\bibfnamefont {Y.~S.}\ \bibnamefont {Zhang}}, \bibinfo {author}
  {\bibfnamefont {J.}~\bibnamefont {Zhu}},\ and\ \bibinfo {author}
  {\bibfnamefont {Y.}~\bibnamefont {Yan}},\ }\bibfield  {title} {\bibinfo
  {title} {Size dependence of {Young’s} modulus in {ZnO} nanowires},\
  }\href@noop {} {\bibfield  {journal} {\bibinfo  {journal} {Physical Review
  Letters}\ }\textbf {\bibinfo {volume} {96}},\ \bibinfo {pages} {075505}
  (\bibinfo {year} {2006})}\BibitemShut {NoStop}%
\bibitem [{\citenamefont {Agrawal}\ \emph {et~al.}(2008)\citenamefont
  {Agrawal}, \citenamefont {Peng}, \citenamefont {Gdoutos},\ and\ \citenamefont
  {Espinosa}}]{agrawal2008elasticity}%
  \BibitemOpen
  \bibfield  {author} {\bibinfo {author} {\bibfnamefont {R.}~\bibnamefont
  {Agrawal}}, \bibinfo {author} {\bibfnamefont {B.}~\bibnamefont {Peng}},
  \bibinfo {author} {\bibfnamefont {E.~E.}\ \bibnamefont {Gdoutos}},\ and\
  \bibinfo {author} {\bibfnamefont {H.~D.}\ \bibnamefont {Espinosa}},\
  }\bibfield  {title} {\bibinfo {title} {Elasticity size effects in {ZnO}
  nanowires- a combined experimental-computational approach},\ }\href@noop {}
  {\bibfield  {journal} {\bibinfo  {journal} {Nano Letters}\ }\textbf {\bibinfo
  {volume} {8}},\ \bibinfo {pages} {3668} (\bibinfo {year} {2008})}\BibitemShut
  {NoStop}%
\bibitem [{\citenamefont {Yang}\ \emph {et~al.}(2012)\citenamefont {Yang},
  \citenamefont {Guo}, \citenamefont {Wang}, \citenamefont {Wang},
  \citenamefont {Qi},\ and\ \citenamefont {Zhang}}]{yang2012size}%
  \BibitemOpen
  \bibfield  {author} {\bibinfo {author} {\bibfnamefont {Y.}~\bibnamefont
  {Yang}}, \bibinfo {author} {\bibfnamefont {W.}~\bibnamefont {Guo}}, \bibinfo
  {author} {\bibfnamefont {X.}~\bibnamefont {Wang}}, \bibinfo {author}
  {\bibfnamefont {Z.}~\bibnamefont {Wang}}, \bibinfo {author} {\bibfnamefont
  {J.}~\bibnamefont {Qi}},\ and\ \bibinfo {author} {\bibfnamefont
  {Y.}~\bibnamefont {Zhang}},\ }\bibfield  {title} {\bibinfo {title} {Size
  dependence of dielectric constant in a single pencil-like {ZnO} nanowire},\
  }\href@noop {} {\bibfield  {journal} {\bibinfo  {journal} {Nano Letters}\
  }\textbf {\bibinfo {volume} {12}},\ \bibinfo {pages} {1919} (\bibinfo {year}
  {2012})}\BibitemShut {NoStop}%
\bibitem [{\citenamefont {Koga}\ \emph {et~al.}(2004)\citenamefont {Koga},
  \citenamefont {Ikeshoji},\ and\ \citenamefont {Sugawara}}]{koga2004size}%
  \BibitemOpen
  \bibfield  {author} {\bibinfo {author} {\bibfnamefont {K.}~\bibnamefont
  {Koga}}, \bibinfo {author} {\bibfnamefont {T.}~\bibnamefont {Ikeshoji}},\
  and\ \bibinfo {author} {\bibfnamefont {K.-i.}\ \bibnamefont {Sugawara}},\
  }\bibfield  {title} {\bibinfo {title} {Size-and temperature-dependent
  structural transitions in gold nanoparticles},\ }\href@noop {} {\bibfield
  {journal} {\bibinfo  {journal} {Physical Review Letters}\ }\textbf {\bibinfo
  {volume} {92}},\ \bibinfo {pages} {115507} (\bibinfo {year}
  {2004})}\BibitemShut {NoStop}%
\bibitem [{\citenamefont {Anderson}\ and\ \citenamefont
  {Lekkerkerker}(2002)}]{anderson2002insights}%
  \BibitemOpen
  \bibfield  {author} {\bibinfo {author} {\bibfnamefont {V.~J.}\ \bibnamefont
  {Anderson}}\ and\ \bibinfo {author} {\bibfnamefont {H.~N.}\ \bibnamefont
  {Lekkerkerker}},\ }\bibfield  {title} {\bibinfo {title} {Insights into phase
  transition kinetics from colloid science},\ }\href@noop {} {\bibfield
  {journal} {\bibinfo  {journal} {Nature}\ }\textbf {\bibinfo {volume} {416}},\
  \bibinfo {pages} {811} (\bibinfo {year} {2002})}\BibitemShut {NoStop}%
\bibitem [{\citenamefont {Poon}(2004)}]{poon2004colloids}%
  \BibitemOpen
  \bibfield  {author} {\bibinfo {author} {\bibfnamefont {W.}~\bibnamefont
  {Poon}},\ }\bibfield  {title} {\bibinfo {title} {Colloids as big atoms},\
  }\href@noop {} {\bibfield  {journal} {\bibinfo  {journal} {Science}\ }\textbf
  {\bibinfo {volume} {304}},\ \bibinfo {pages} {830} (\bibinfo {year}
  {2004})}\BibitemShut {NoStop}%
\bibitem [{\citenamefont {Manoharan}(2015)}]{manoharan2015colloidal}%
  \BibitemOpen
  \bibfield  {author} {\bibinfo {author} {\bibfnamefont {V.~N.}\ \bibnamefont
  {Manoharan}},\ }\bibfield  {title} {\bibinfo {title} {Colloidal matter:
  Packing, geometry, and entropy},\ }\href@noop {} {\bibfield  {journal}
  {\bibinfo  {journal} {Science}\ }\textbf {\bibinfo {volume} {349}} (\bibinfo
  {year} {2015})}\BibitemShut {NoStop}%
\bibitem [{\citenamefont {Rubin-Zuzic}\ \emph {et~al.}(2006)\citenamefont
  {Rubin-Zuzic}, \citenamefont {Morfill}, \citenamefont {Ivlev}, \citenamefont
  {Pompl}, \citenamefont {Klumov}, \citenamefont {Bunk}, \citenamefont
  {Thomas}, \citenamefont {Rothermel}, \citenamefont {Havnes},\ and\
  \citenamefont {Fouquet}}]{rubin2006kinetic}%
  \BibitemOpen
  \bibfield  {author} {\bibinfo {author} {\bibfnamefont {M.}~\bibnamefont
  {Rubin-Zuzic}}, \bibinfo {author} {\bibfnamefont {G.}~\bibnamefont
  {Morfill}}, \bibinfo {author} {\bibfnamefont {A.}~\bibnamefont {Ivlev}},
  \bibinfo {author} {\bibfnamefont {R.}~\bibnamefont {Pompl}}, \bibinfo
  {author} {\bibfnamefont {B.}~\bibnamefont {Klumov}}, \bibinfo {author}
  {\bibfnamefont {W.}~\bibnamefont {Bunk}}, \bibinfo {author} {\bibfnamefont
  {H.}~\bibnamefont {Thomas}}, \bibinfo {author} {\bibfnamefont
  {H.}~\bibnamefont {Rothermel}}, \bibinfo {author} {\bibfnamefont
  {O.}~\bibnamefont {Havnes}},\ and\ \bibinfo {author} {\bibfnamefont
  {A.}~\bibnamefont {Fouquet}},\ }\bibfield  {title} {\bibinfo {title} {Kinetic
  development of crystallization fronts in complex plasmas},\ }\href@noop {}
  {\bibfield  {journal} {\bibinfo  {journal} {Nature Physics}\ }\textbf
  {\bibinfo {volume} {2}},\ \bibinfo {pages} {181} (\bibinfo {year}
  {2006})}\BibitemShut {NoStop}%
\bibitem [{\citenamefont {Knapek}\ \emph {et~al.}(2007)\citenamefont {Knapek},
  \citenamefont {Samsonov}, \citenamefont {Zhdanov}, \citenamefont {Konopka},\
  and\ \citenamefont {Morfill}}]{knapek2007recrystallization}%
  \BibitemOpen
  \bibfield  {author} {\bibinfo {author} {\bibfnamefont {C.}~\bibnamefont
  {Knapek}}, \bibinfo {author} {\bibfnamefont {D.}~\bibnamefont {Samsonov}},
  \bibinfo {author} {\bibfnamefont {S.}~\bibnamefont {Zhdanov}}, \bibinfo
  {author} {\bibfnamefont {U.}~\bibnamefont {Konopka}},\ and\ \bibinfo {author}
  {\bibfnamefont {G.}~\bibnamefont {Morfill}},\ }\bibfield  {title} {\bibinfo
  {title} {Recrystallization of a {2D} plasma crystal},\ }\href@noop {}
  {\bibfield  {journal} {\bibinfo  {journal} {Physical Review Letters}\
  }\textbf {\bibinfo {volume} {98}},\ \bibinfo {pages} {015004} (\bibinfo
  {year} {2007})}\BibitemShut {NoStop}%
\bibitem [{\citenamefont {S{\"u}tterlin}\ \emph {et~al.}(2009)\citenamefont
  {S{\"u}tterlin}, \citenamefont {Wysocki}, \citenamefont {Ivlev},
  \citenamefont {R{\"a}th}, \citenamefont {Thomas}, \citenamefont
  {Rubin-Zuzic}, \citenamefont {Goedheer}, \citenamefont {Fortov},
  \citenamefont {Lipaev}, \citenamefont {Molotkov} \emph
  {et~al.}}]{sutterlin2009dynamics}%
  \BibitemOpen
  \bibfield  {author} {\bibinfo {author} {\bibfnamefont {K.}~\bibnamefont
  {S{\"u}tterlin}}, \bibinfo {author} {\bibfnamefont {A.}~\bibnamefont
  {Wysocki}}, \bibinfo {author} {\bibfnamefont {A.}~\bibnamefont {Ivlev}},
  \bibinfo {author} {\bibfnamefont {C.}~\bibnamefont {R{\"a}th}}, \bibinfo
  {author} {\bibfnamefont {H.}~\bibnamefont {Thomas}}, \bibinfo {author}
  {\bibfnamefont {M.}~\bibnamefont {Rubin-Zuzic}}, \bibinfo {author}
  {\bibfnamefont {W.}~\bibnamefont {Goedheer}}, \bibinfo {author}
  {\bibfnamefont {V.}~\bibnamefont {Fortov}}, \bibinfo {author} {\bibfnamefont
  {A.}~\bibnamefont {Lipaev}}, \bibinfo {author} {\bibfnamefont
  {V.}~\bibnamefont {Molotkov}}, \emph {et~al.},\ }\bibfield  {title} {\bibinfo
  {title} {Dynamics of lane formation in driven binary complex plasmas},\
  }\href@noop {} {\bibfield  {journal} {\bibinfo  {journal} {Physical Review
  Letters}\ }\textbf {\bibinfo {volume} {102}},\ \bibinfo {pages} {085003}
  (\bibinfo {year} {2009})}\BibitemShut {NoStop}%
\bibitem [{\citenamefont {Lim}\ \emph {et~al.}(2019{\natexlab{a}})\citenamefont
  {Lim}, \citenamefont {Souslov}, \citenamefont {Vitelli},\ and\ \citenamefont
  {Jaeger}}]{lim2019cluster}%
  \BibitemOpen
  \bibfield  {author} {\bibinfo {author} {\bibfnamefont {M.~X.}\ \bibnamefont
  {Lim}}, \bibinfo {author} {\bibfnamefont {A.}~\bibnamefont {Souslov}},
  \bibinfo {author} {\bibfnamefont {V.}~\bibnamefont {Vitelli}},\ and\ \bibinfo
  {author} {\bibfnamefont {H.~M.}\ \bibnamefont {Jaeger}},\ }\bibfield  {title}
  {\bibinfo {title} {Cluster formation by acoustic forces and active
  fluctuations in levitated granular matter},\ }\href@noop {} {\bibfield
  {journal} {\bibinfo  {journal} {Nature Physics}\ }\textbf {\bibinfo {volume}
  {15}},\ \bibinfo {pages} {460} (\bibinfo {year}
  {2019}{\natexlab{a}})}\BibitemShut {NoStop}%
\bibitem [{\citenamefont {Lim}\ \emph {et~al.}(2019{\natexlab{b}})\citenamefont
  {Lim}, \citenamefont {Murphy},\ and\ \citenamefont {Jaeger}}]{lim2019edges}%
  \BibitemOpen
  \bibfield  {author} {\bibinfo {author} {\bibfnamefont {M.~X.}\ \bibnamefont
  {Lim}}, \bibinfo {author} {\bibfnamefont {K.~A.}\ \bibnamefont {Murphy}},\
  and\ \bibinfo {author} {\bibfnamefont {H.~M.}\ \bibnamefont {Jaeger}},\
  }\bibfield  {title} {\bibinfo {title} {Edges control clustering in levitated
  granular matter},\ }\href@noop {} {\bibfield  {journal} {\bibinfo  {journal}
  {Granular Matter}\ }\textbf {\bibinfo {volume} {21}},\ \bibinfo {pages} {77}
  (\bibinfo {year} {2019}{\natexlab{b}})}\BibitemShut {NoStop}%
\bibitem [{\citenamefont {Silva}\ and\ \citenamefont
  {Bruus}(2014)}]{silva2014acoustic}%
  \BibitemOpen
  \bibfield  {author} {\bibinfo {author} {\bibfnamefont {G.~T.}\ \bibnamefont
  {Silva}}\ and\ \bibinfo {author} {\bibfnamefont {H.}~\bibnamefont {Bruus}},\
  }\bibfield  {title} {\bibinfo {title} {Acoustic interaction forces between
  small particles in an ideal fluid},\ }\href@noop {} {\bibfield  {journal}
  {\bibinfo  {journal} {Physical Review E}\ }\textbf {\bibinfo {volume} {90}},\
  \bibinfo {pages} {063007} (\bibinfo {year} {2014})}\BibitemShut {NoStop}%
\bibitem [{\citenamefont {Sepehrirahnama}\ \emph {et~al.}(2015)\citenamefont
  {Sepehrirahnama}, \citenamefont {Lim},\ and\ \citenamefont
  {Chau}}]{sepehrirahnama2015numerical}%
  \BibitemOpen
  \bibfield  {author} {\bibinfo {author} {\bibfnamefont {S.}~\bibnamefont
  {Sepehrirahnama}}, \bibinfo {author} {\bibfnamefont {K.-M.}\ \bibnamefont
  {Lim}},\ and\ \bibinfo {author} {\bibfnamefont {F.~S.}\ \bibnamefont
  {Chau}},\ }\bibfield  {title} {\bibinfo {title} {Numerical study of
  interparticle radiation force acting on rigid spheres in a standing wave},\
  }\href@noop {} {\bibfield  {journal} {\bibinfo  {journal} {The Journal of the
  Acoustical Society of America}\ }\textbf {\bibinfo {volume} {137}},\ \bibinfo
  {pages} {2614} (\bibinfo {year} {2015})}\BibitemShut {NoStop}%
\bibitem [{\citenamefont {Tsai}\ \emph {et~al.}(2005)\citenamefont {Tsai},
  \citenamefont {Ye}, \citenamefont {Rodriguez}, \citenamefont {Gollub},\ and\
  \citenamefont {Lubensky}}]{tsai2005chiral}%
  \BibitemOpen
  \bibfield  {author} {\bibinfo {author} {\bibfnamefont {J.-C.}\ \bibnamefont
  {Tsai}}, \bibinfo {author} {\bibfnamefont {F.}~\bibnamefont {Ye}}, \bibinfo
  {author} {\bibfnamefont {J.}~\bibnamefont {Rodriguez}}, \bibinfo {author}
  {\bibfnamefont {J.~P.}\ \bibnamefont {Gollub}},\ and\ \bibinfo {author}
  {\bibfnamefont {T.}~\bibnamefont {Lubensky}},\ }\bibfield  {title} {\bibinfo
  {title} {A chiral granular gas},\ }\href@noop {} {\bibfield  {journal}
  {\bibinfo  {journal} {Physical Review Letters}\ }\textbf {\bibinfo {volume}
  {94}},\ \bibinfo {pages} {214301} (\bibinfo {year} {2005})}\BibitemShut
  {NoStop}%
\bibitem [{\citenamefont {Briand}\ and\ \citenamefont
  {Dauchot}(2016)}]{briand2016crystallization}%
  \BibitemOpen
  \bibfield  {author} {\bibinfo {author} {\bibfnamefont {G.}~\bibnamefont
  {Briand}}\ and\ \bibinfo {author} {\bibfnamefont {O.}~\bibnamefont
  {Dauchot}},\ }\bibfield  {title} {\bibinfo {title} {Crystallization of
  self-propelled hard discs},\ }\href@noop {} {\bibfield  {journal} {\bibinfo
  {journal} {Physical Review Letters}\ }\textbf {\bibinfo {volume} {117}},\
  \bibinfo {pages} {098004} (\bibinfo {year} {2016})}\BibitemShut {NoStop}%
\bibitem [{\citenamefont {Scholz}\ \emph {et~al.}(2018)\citenamefont {Scholz},
  \citenamefont {Jahanshahi}, \citenamefont {Ldov},\ and\ \citenamefont
  {L{\"o}wen}}]{scholz2018inertial}%
  \BibitemOpen
  \bibfield  {author} {\bibinfo {author} {\bibfnamefont {C.}~\bibnamefont
  {Scholz}}, \bibinfo {author} {\bibfnamefont {S.}~\bibnamefont {Jahanshahi}},
  \bibinfo {author} {\bibfnamefont {A.}~\bibnamefont {Ldov}},\ and\ \bibinfo
  {author} {\bibfnamefont {H.}~\bibnamefont {L{\"o}wen}},\ }\bibfield  {title}
  {\bibinfo {title} {Inertial delay of self-propelled particles},\ }\href@noop
  {} {\bibfield  {journal} {\bibinfo  {journal} {Nature Communications}\
  }\textbf {\bibinfo {volume} {9}},\ \bibinfo {pages} {1} (\bibinfo {year}
  {2018})}\BibitemShut {NoStop}%
\bibitem [{\citenamefont {Thoroddsen}\ \emph {et~al.}(2005)\citenamefont
  {Thoroddsen}, \citenamefont {Takehara},\ and\ \citenamefont
  {Etoh}}]{thoroddsen2005coalescence}%
  \BibitemOpen
  \bibfield  {author} {\bibinfo {author} {\bibfnamefont {S.}~\bibnamefont
  {Thoroddsen}}, \bibinfo {author} {\bibfnamefont {K.}~\bibnamefont
  {Takehara}},\ and\ \bibinfo {author} {\bibfnamefont {T.}~\bibnamefont
  {Etoh}},\ }\bibfield  {title} {\bibinfo {title} {The coalescence speed of a
  pendent and a sessile drop},\ }\href@noop {} {\bibfield  {journal} {\bibinfo
  {journal} {Journal of Fluid Mechanics}\ }\textbf {\bibinfo {volume} {527}},\
  \bibinfo {pages} {85} (\bibinfo {year} {2005})}\BibitemShut {NoStop}%
\bibitem [{\citenamefont {Aarts}\ \emph {et~al.}(2005)\citenamefont {Aarts},
  \citenamefont {Lekkerkerker}, \citenamefont {Guo}, \citenamefont {Wegdam},\
  and\ \citenamefont {Bonn}}]{aarts2005hydrodynamics}%
  \BibitemOpen
  \bibfield  {author} {\bibinfo {author} {\bibfnamefont {D.~G.}\ \bibnamefont
  {Aarts}}, \bibinfo {author} {\bibfnamefont {H.~N.}\ \bibnamefont
  {Lekkerkerker}}, \bibinfo {author} {\bibfnamefont {H.}~\bibnamefont {Guo}},
  \bibinfo {author} {\bibfnamefont {G.~H.}\ \bibnamefont {Wegdam}},\ and\
  \bibinfo {author} {\bibfnamefont {D.}~\bibnamefont {Bonn}},\ }\bibfield
  {title} {\bibinfo {title} {Hydrodynamics of droplet coalescence},\
  }\href@noop {} {\bibfield  {journal} {\bibinfo  {journal} {Physical Review
  Letters}\ }\textbf {\bibinfo {volume} {95}},\ \bibinfo {pages} {164503}
  (\bibinfo {year} {2005})}\BibitemShut {NoStop}%
\bibitem [{\citenamefont {Hertlein}\ \emph {et~al.}(2008)\citenamefont
  {Hertlein}, \citenamefont {Helden}, \citenamefont {Gambassi}, \citenamefont
  {Dietrich},\ and\ \citenamefont {Bechinger}}]{hertlein2008direct}%
  \BibitemOpen
  \bibfield  {author} {\bibinfo {author} {\bibfnamefont {C.}~\bibnamefont
  {Hertlein}}, \bibinfo {author} {\bibfnamefont {L.}~\bibnamefont {Helden}},
  \bibinfo {author} {\bibfnamefont {A.}~\bibnamefont {Gambassi}}, \bibinfo
  {author} {\bibfnamefont {S.}~\bibnamefont {Dietrich}},\ and\ \bibinfo
  {author} {\bibfnamefont {C.}~\bibnamefont {Bechinger}},\ }\bibfield  {title}
  {\bibinfo {title} {Direct measurement of critical casimir forces},\
  }\href@noop {} {\bibfield  {journal} {\bibinfo  {journal} {Nature}\ }\textbf
  {\bibinfo {volume} {451}},\ \bibinfo {pages} {172} (\bibinfo {year}
  {2008})}\BibitemShut {NoStop}%
\bibitem [{\citenamefont {Yu}\ \emph {et~al.}(2003)\citenamefont {Yu},
  \citenamefont {Mei}, \citenamefont {Luo},\ and\ \citenamefont
  {Shyy}}]{yu2003visc}%
  \BibitemOpen
  \bibfield  {author} {\bibinfo {author} {\bibfnamefont {D.}~\bibnamefont
  {Yu}}, \bibinfo {author} {\bibfnamefont {R.}~\bibnamefont {Mei}}, \bibinfo
  {author} {\bibfnamefont {L.-S.}\ \bibnamefont {Luo}},\ and\ \bibinfo {author}
  {\bibfnamefont {W.}~\bibnamefont {Shyy}},\ }\bibfield  {title} {\bibinfo
  {title} {Viscous flow computations with the method of lattice {Boltzmann}
  equation},\ }\href {https://doi.org/10.1016/S0376-0421(03)00003-4} {\bibfield
   {journal} {\bibinfo  {journal} {Progress in Aerospace Sciences}\ }\textbf
  {\bibinfo {volume} {39}},\ \bibinfo {pages} {329} (\bibinfo {year}
  {2003})}\BibitemShut {NoStop}%
\bibitem [{\citenamefont {Bauer}\ \emph {et~al.}(2021)\citenamefont {Bauer},
  \citenamefont {Eibl}, \citenamefont {Godenschwager}, \citenamefont {Kohl},
  \citenamefont {Kuron}, \citenamefont {Rettinger}, \citenamefont {Schornbaum},
  \citenamefont {Schwarzmeier}, \citenamefont {Th{\"o}nnes}, \citenamefont
  {K{\"o}stler} \emph {et~al.}}]{bauer2020walberla}%
  \BibitemOpen
  \bibfield  {author} {\bibinfo {author} {\bibfnamefont {M.}~\bibnamefont
  {Bauer}}, \bibinfo {author} {\bibfnamefont {S.}~\bibnamefont {Eibl}},
  \bibinfo {author} {\bibfnamefont {C.}~\bibnamefont {Godenschwager}}, \bibinfo
  {author} {\bibfnamefont {N.}~\bibnamefont {Kohl}}, \bibinfo {author}
  {\bibfnamefont {M.}~\bibnamefont {Kuron}}, \bibinfo {author} {\bibfnamefont
  {C.}~\bibnamefont {Rettinger}}, \bibinfo {author} {\bibfnamefont
  {F.}~\bibnamefont {Schornbaum}}, \bibinfo {author} {\bibfnamefont
  {C.}~\bibnamefont {Schwarzmeier}}, \bibinfo {author} {\bibfnamefont
  {D.}~\bibnamefont {Th{\"o}nnes}}, \bibinfo {author} {\bibfnamefont
  {H.}~\bibnamefont {K{\"o}stler}}, \emph {et~al.},\ }\bibfield  {title}
  {\bibinfo {title} {{waLBerla}: A block-structured high-performance framework
  for multiphysics simulations},\ }\href@noop {} {\bibfield  {journal}
  {\bibinfo  {journal} {Computers \& Mathematics with Applications}\ }\textbf
  {\bibinfo {volume} {81}},\ \bibinfo {pages} {478} (\bibinfo {year}
  {2021})}\BibitemShut {NoStop}%
\bibitem [{\citenamefont {G{\"o}tz}\ \emph {et~al.}(2010)\citenamefont
  {G{\"o}tz}, \citenamefont {Iglberger}, \citenamefont {Feichtinger},
  \citenamefont {Donath},\ and\ \citenamefont {R{\"u}de}}]{gotz2010pe}%
  \BibitemOpen
  \bibfield  {author} {\bibinfo {author} {\bibfnamefont {J.}~\bibnamefont
  {G{\"o}tz}}, \bibinfo {author} {\bibfnamefont {K.}~\bibnamefont {Iglberger}},
  \bibinfo {author} {\bibfnamefont {C.}~\bibnamefont {Feichtinger}}, \bibinfo
  {author} {\bibfnamefont {S.}~\bibnamefont {Donath}},\ and\ \bibinfo {author}
  {\bibfnamefont {U.}~\bibnamefont {R{\"u}de}},\ }\bibfield  {title} {\bibinfo
  {title} {Coupling multibody dynamics and computational fluid dynamics on 8192
  processor cores},\ }\href {https://doi.org/10.1016/j.parco.2010.01.005}
  {\bibfield  {journal} {\bibinfo  {journal} {Parallel Computing}\ }\textbf
  {\bibinfo {volume} {36}},\ \bibinfo {pages} {142} (\bibinfo {year}
  {2010})}\BibitemShut {NoStop}%
\bibitem [{\citenamefont {Owen}\ \emph {et~al.}(2011)\citenamefont {Owen},
  \citenamefont {Leonardi},\ and\ \citenamefont {Feng}}]{owen2011efficient}%
  \BibitemOpen
  \bibfield  {author} {\bibinfo {author} {\bibfnamefont {D.}~\bibnamefont
  {Owen}}, \bibinfo {author} {\bibfnamefont {C.}~\bibnamefont {Leonardi}},\
  and\ \bibinfo {author} {\bibfnamefont {Y.}~\bibnamefont {Feng}},\ }\bibfield
  {title} {\bibinfo {title} {An efficient framework for fluid--structure
  interaction using the lattice boltzmann method and immersed moving
  boundaries},\ }\href@noop {} {\bibfield  {journal} {\bibinfo  {journal}
  {International Journal for Numerical Methods in Engineering}\ }\textbf
  {\bibinfo {volume} {87}},\ \bibinfo {pages} {66} (\bibinfo {year}
  {2011})}\BibitemShut {NoStop}%
\bibitem [{\citenamefont {Settnes}\ and\ \citenamefont
  {Bruus}(2012)}]{settnes2012forces}%
  \BibitemOpen
  \bibfield  {author} {\bibinfo {author} {\bibfnamefont {M.}~\bibnamefont
  {Settnes}}\ and\ \bibinfo {author} {\bibfnamefont {H.}~\bibnamefont
  {Bruus}},\ }\bibfield  {title} {\bibinfo {title} {Forces acting on a small
  particle in an acoustical field in a viscous fluid},\ }\href@noop {}
  {\bibfield  {journal} {\bibinfo  {journal} {Physical Review E}\ }\textbf
  {\bibinfo {volume} {85}},\ \bibinfo {pages} {016327} (\bibinfo {year}
  {2012})}\BibitemShut {NoStop}%
\bibitem [{\citenamefont {Rudnick}\ and\ \citenamefont
  {Barmatz}(1990)}]{rudnick1990oscillational}%
  \BibitemOpen
  \bibfield  {author} {\bibinfo {author} {\bibfnamefont {J.}~\bibnamefont
  {Rudnick}}\ and\ \bibinfo {author} {\bibfnamefont {M.}~\bibnamefont
  {Barmatz}},\ }\bibfield  {title} {\bibinfo {title} {Oscillational
  instabilities in single-mode acoustic levitators},\ }\href@noop {} {\bibfield
   {journal} {\bibinfo  {journal} {The Journal of the Acoustical Society of
  America}\ }\textbf {\bibinfo {volume} {87}},\ \bibinfo {pages} {81} (\bibinfo
  {year} {1990})}\BibitemShut {NoStop}%
\bibitem [{\citenamefont {Trinh}\ and\ \citenamefont
  {Robey}(1994)}]{trinh1994experimental}%
  \BibitemOpen
  \bibfield  {author} {\bibinfo {author} {\bibfnamefont {E.}~\bibnamefont
  {Trinh}}\ and\ \bibinfo {author} {\bibfnamefont {J.}~\bibnamefont {Robey}},\
  }\bibfield  {title} {\bibinfo {title} {Experimental study of streaming flows
  associated with ultrasonic levitators},\ }\href@noop {} {\bibfield  {journal}
  {\bibinfo  {journal} {Physics of Fluids}\ }\textbf {\bibinfo {volume} {6}},\
  \bibinfo {pages} {3567} (\bibinfo {year} {1994})}\BibitemShut {NoStop}%
\bibitem [{\citenamefont {Baer}\ \emph {et~al.}(2011)\citenamefont {Baer},
  \citenamefont {Andrade}, \citenamefont {Esen}, \citenamefont {Adamowski},
  \citenamefont {Schweiger},\ and\ \citenamefont
  {Ostendorf}}]{baer2011analysis}%
  \BibitemOpen
  \bibfield  {author} {\bibinfo {author} {\bibfnamefont {S.}~\bibnamefont
  {Baer}}, \bibinfo {author} {\bibfnamefont {M.~A.}\ \bibnamefont {Andrade}},
  \bibinfo {author} {\bibfnamefont {C.}~\bibnamefont {Esen}}, \bibinfo {author}
  {\bibfnamefont {J.~C.}\ \bibnamefont {Adamowski}}, \bibinfo {author}
  {\bibfnamefont {G.}~\bibnamefont {Schweiger}},\ and\ \bibinfo {author}
  {\bibfnamefont {A.}~\bibnamefont {Ostendorf}},\ }\bibfield  {title} {\bibinfo
  {title} {Analysis of the particle stability in a new designed ultrasonic
  levitation device},\ }\href@noop {} {\bibfield  {journal} {\bibinfo
  {journal} {Review of Scientific Instruments}\ }\textbf {\bibinfo {volume}
  {82}},\ \bibinfo {pages} {105111} (\bibinfo {year} {2011})}\BibitemShut
  {NoStop}%
\bibitem [{\citenamefont {Cammarata}\ and\ \citenamefont
  {Sieradzki}(1994)}]{cammarata1994surface}%
  \BibitemOpen
  \bibfield  {author} {\bibinfo {author} {\bibfnamefont {R.~C.}\ \bibnamefont
  {Cammarata}}\ and\ \bibinfo {author} {\bibfnamefont {K.}~\bibnamefont
  {Sieradzki}},\ }\bibfield  {title} {\bibinfo {title} {Surface and interface
  stresses},\ }\href@noop {} {\bibfield  {journal} {\bibinfo  {journal} {Annual
  Review of Materials Science}\ }\textbf {\bibinfo {volume} {24}},\ \bibinfo
  {pages} {215} (\bibinfo {year} {1994})}\BibitemShut {NoStop}%
\bibitem [{\citenamefont {Haiss}(2001)}]{haiss2001surface}%
  \BibitemOpen
  \bibfield  {author} {\bibinfo {author} {\bibfnamefont {W.}~\bibnamefont
  {Haiss}},\ }\bibfield  {title} {\bibinfo {title} {Surface stress of clean and
  adsorbate-covered solids},\ }\href@noop {} {\bibfield  {journal} {\bibinfo
  {journal} {Reports on Progress in Physics}\ }\textbf {\bibinfo {volume}
  {64}},\ \bibinfo {pages} {591} (\bibinfo {year} {2001})}\BibitemShut
  {NoStop}%
\bibitem [{\citenamefont {Trinh}\ \emph {et~al.}(1988)\citenamefont {Trinh},
  \citenamefont {Marston},\ and\ \citenamefont {Robey}}]{trinh1988acoustic}%
  \BibitemOpen
  \bibfield  {author} {\bibinfo {author} {\bibfnamefont {E.}~\bibnamefont
  {Trinh}}, \bibinfo {author} {\bibfnamefont {P.}~\bibnamefont {Marston}},\
  and\ \bibinfo {author} {\bibfnamefont {J.}~\bibnamefont {Robey}},\ }\bibfield
   {title} {\bibinfo {title} {Acoustic measurement of the surface tension of
  levitated drops},\ }\href@noop {} {\bibfield  {journal} {\bibinfo  {journal}
  {Journal of Colloid and Interface Science}\ }\textbf {\bibinfo {volume}
  {124}},\ \bibinfo {pages} {95} (\bibinfo {year} {1988})}\BibitemShut
  {NoStop}%
\bibitem [{\citenamefont {Tian}\ \emph {et~al.}(1995)\citenamefont {Tian},
  \citenamefont {Holt},\ and\ \citenamefont {Apfel}}]{tian1995new}%
  \BibitemOpen
  \bibfield  {author} {\bibinfo {author} {\bibfnamefont {Y.}~\bibnamefont
  {Tian}}, \bibinfo {author} {\bibfnamefont {R.~G.}\ \bibnamefont {Holt}},\
  and\ \bibinfo {author} {\bibfnamefont {R.~E.}\ \bibnamefont {Apfel}},\
  }\bibfield  {title} {\bibinfo {title} {A new method for measuring liquid
  surface tension with acoustic levitation},\ }\href@noop {} {\bibfield
  {journal} {\bibinfo  {journal} {Review of scientific instruments}\ }\textbf
  {\bibinfo {volume} {66}},\ \bibinfo {pages} {3349} (\bibinfo {year}
  {1995})}\BibitemShut {NoStop}%
\bibitem [{\citenamefont {Wang}\ \emph {et~al.}(1994)\citenamefont {Wang},
  \citenamefont {Anilkumar}, \citenamefont {Lee},\ and\ \citenamefont
  {Lin}}]{wang1994bifurcation}%
  \BibitemOpen
  \bibfield  {author} {\bibinfo {author} {\bibfnamefont {T.~G.}\ \bibnamefont
  {Wang}}, \bibinfo {author} {\bibfnamefont {A.}~\bibnamefont {Anilkumar}},
  \bibinfo {author} {\bibfnamefont {C.}~\bibnamefont {Lee}},\ and\ \bibinfo
  {author} {\bibfnamefont {K.}~\bibnamefont {Lin}},\ }\bibfield  {title}
  {\bibinfo {title} {Bifurcation of rotating liquid drops: results from
  {USML}-1 experiments in space},\ }\href@noop {} {\bibfield  {journal}
  {\bibinfo  {journal} {Journal of Fluid Mechanics}\ }\textbf {\bibinfo
  {volume} {276}},\ \bibinfo {pages} {389} (\bibinfo {year}
  {1994})}\BibitemShut {NoStop}%
\bibitem [{\citenamefont {Lewis}\ \emph {et~al.}(1987)\citenamefont {Lewis},
  \citenamefont {Marsden},\ and\ \citenamefont {Ratiu}}]{lewis1987stability}%
  \BibitemOpen
  \bibfield  {author} {\bibinfo {author} {\bibfnamefont {D.}~\bibnamefont
  {Lewis}}, \bibinfo {author} {\bibfnamefont {J.}~\bibnamefont {Marsden}},\
  and\ \bibinfo {author} {\bibfnamefont {T.}~\bibnamefont {Ratiu}},\ }\bibfield
   {title} {\bibinfo {title} {Stability and bifurcation of a rotating planar
  liquid drop},\ }\href@noop {} {\bibfield  {journal} {\bibinfo  {journal}
  {Journal of Mathematical Physics}\ }\textbf {\bibinfo {volume} {28}},\
  \bibinfo {pages} {2508} (\bibinfo {year} {1987})}\BibitemShut {NoStop}%
\bibitem [{\citenamefont {Nguyen}\ \emph {et~al.}(2018)\citenamefont {Nguyen},
  \citenamefont {Schoemaker}, \citenamefont {Blokhuis},\ and\ \citenamefont
  {Schall}}]{nguyen2018measurement}%
  \BibitemOpen
  \bibfield  {author} {\bibinfo {author} {\bibfnamefont {V.}~\bibnamefont
  {Nguyen}}, \bibinfo {author} {\bibfnamefont {F.}~\bibnamefont {Schoemaker}},
  \bibinfo {author} {\bibfnamefont {E.}~\bibnamefont {Blokhuis}},\ and\
  \bibinfo {author} {\bibfnamefont {P.}~\bibnamefont {Schall}},\ }\bibfield
  {title} {\bibinfo {title} {Measurement of the curvature-dependent surface
  tension in nucleating colloidal liquids},\ }\href@noop {} {\bibfield
  {journal} {\bibinfo  {journal} {Physical Review Letters}\ }\textbf {\bibinfo
  {volume} {121}},\ \bibinfo {pages} {246102} (\bibinfo {year}
  {2018})}\BibitemShut {NoStop}%
\bibitem [{\citenamefont {Lau}\ \emph {et~al.}(2015)\citenamefont {Lau},
  \citenamefont {Hunt}, \citenamefont {M{\"u}ller}, \citenamefont {Jackson},\
  and\ \citenamefont {Ford}}]{lau2015water}%
  \BibitemOpen
  \bibfield  {author} {\bibinfo {author} {\bibfnamefont {G.~V.}\ \bibnamefont
  {Lau}}, \bibinfo {author} {\bibfnamefont {P.~A.}\ \bibnamefont {Hunt}},
  \bibinfo {author} {\bibfnamefont {E.~A.}\ \bibnamefont {M{\"u}ller}},
  \bibinfo {author} {\bibfnamefont {G.}~\bibnamefont {Jackson}},\ and\ \bibinfo
  {author} {\bibfnamefont {I.~J.}\ \bibnamefont {Ford}},\ }\bibfield  {title}
  {\bibinfo {title} {Water droplet excess free energy determined by cluster
  mitosis using guided molecular dynamics},\ }\href@noop {} {\bibfield
  {journal} {\bibinfo  {journal} {The Journal of Chemical Physics}\ }\textbf
  {\bibinfo {volume} {143}},\ \bibinfo {pages} {244709} (\bibinfo {year}
  {2015})}\BibitemShut {NoStop}%
\bibitem [{\citenamefont {Zhang}\ \emph {et~al.}(2016)\citenamefont {Zhang},
  \citenamefont {Qiu}, \citenamefont {Wang}, \citenamefont {Ke},\ and\
  \citenamefont {Liu}}]{zhang2016acoustically}%
  \BibitemOpen
  \bibfield  {author} {\bibinfo {author} {\bibfnamefont {S.}~\bibnamefont
  {Zhang}}, \bibinfo {author} {\bibfnamefont {C.}~\bibnamefont {Qiu}}, \bibinfo
  {author} {\bibfnamefont {M.}~\bibnamefont {Wang}}, \bibinfo {author}
  {\bibfnamefont {M.}~\bibnamefont {Ke}},\ and\ \bibinfo {author}
  {\bibfnamefont {Z.}~\bibnamefont {Liu}},\ }\bibfield  {title} {\bibinfo
  {title} {Acoustically mediated long-range interaction among multiple
  spherical particles exposed to a plane standing wave},\ }\href@noop {}
  {\bibfield  {journal} {\bibinfo  {journal} {New Journal of Physics}\ }\textbf
  {\bibinfo {volume} {18}},\ \bibinfo {pages} {113034} (\bibinfo {year}
  {2016})}\BibitemShut {NoStop}%
\bibitem [{\citenamefont {Schall}\ \emph {et~al.}(2004)\citenamefont {Schall},
  \citenamefont {Cohen}, \citenamefont {Weitz},\ and\ \citenamefont
  {Spaepen}}]{schall2004visualization}%
  \BibitemOpen
  \bibfield  {author} {\bibinfo {author} {\bibfnamefont {P.}~\bibnamefont
  {Schall}}, \bibinfo {author} {\bibfnamefont {I.}~\bibnamefont {Cohen}},
  \bibinfo {author} {\bibfnamefont {D.~A.}\ \bibnamefont {Weitz}},\ and\
  \bibinfo {author} {\bibfnamefont {F.}~\bibnamefont {Spaepen}},\ }\bibfield
  {title} {\bibinfo {title} {Visualization of dislocation dynamics in colloidal
  crystals},\ }\href@noop {} {\bibfield  {journal} {\bibinfo  {journal}
  {Science}\ }\textbf {\bibinfo {volume} {305}},\ \bibinfo {pages} {1944}
  (\bibinfo {year} {2004})}\BibitemShut {NoStop}%
\bibitem [{\citenamefont {Suenaga}\ \emph {et~al.}(2007)\citenamefont
  {Suenaga}, \citenamefont {Wakabayashi}, \citenamefont {Koshino},
  \citenamefont {Sato}, \citenamefont {Urita},\ and\ \citenamefont
  {Iijima}}]{suenaga2007imaging}%
  \BibitemOpen
  \bibfield  {author} {\bibinfo {author} {\bibfnamefont {K.}~\bibnamefont
  {Suenaga}}, \bibinfo {author} {\bibfnamefont {H.}~\bibnamefont
  {Wakabayashi}}, \bibinfo {author} {\bibfnamefont {M.}~\bibnamefont
  {Koshino}}, \bibinfo {author} {\bibfnamefont {Y.}~\bibnamefont {Sato}},
  \bibinfo {author} {\bibfnamefont {K.}~\bibnamefont {Urita}},\ and\ \bibinfo
  {author} {\bibfnamefont {S.}~\bibnamefont {Iijima}},\ }\bibfield  {title}
  {\bibinfo {title} {Imaging active topological defects in carbon nanotubes},\
  }\href@noop {} {\bibfield  {journal} {\bibinfo  {journal} {Nature
  Nanotechnology}\ }\textbf {\bibinfo {volume} {2}},\ \bibinfo {pages} {358}
  (\bibinfo {year} {2007})}\BibitemShut {NoStop}%
\bibitem [{\citenamefont {Chaikin}\ and\ \citenamefont
  {Lubensky}()}]{chaikin1995principles}%
  \BibitemOpen
  \bibfield  {author} {\bibinfo {author} {\bibfnamefont {P.~M.}\ \bibnamefont
  {Chaikin}}\ and\ \bibinfo {author} {\bibfnamefont {T.~C.}\ \bibnamefont
  {Lubensky}},\ }\href@noop {} {\emph {\bibinfo {title} {Principles of
  condensed matter physics}}},\ Vol.~\bibinfo {volume} {10}\BibitemShut
  {NoStop}%
\bibitem [{\citenamefont {Greer}\ and\ \citenamefont
  {Nix}(2006)}]{greer2006nanoscale}%
  \BibitemOpen
  \bibfield  {author} {\bibinfo {author} {\bibfnamefont {J.~R.}\ \bibnamefont
  {Greer}}\ and\ \bibinfo {author} {\bibfnamefont {W.~D.}\ \bibnamefont
  {Nix}},\ }\bibfield  {title} {\bibinfo {title} {Nanoscale gold pillars
  strengthened through dislocation starvation},\ }\href@noop {} {\bibfield
  {journal} {\bibinfo  {journal} {Physical Review B}\ }\textbf {\bibinfo
  {volume} {73}},\ \bibinfo {pages} {245410} (\bibinfo {year}
  {2006})}\BibitemShut {NoStop}%
\bibitem [{\citenamefont {Jang}\ and\ \citenamefont
  {Greer}(2010)}]{jang2010transition}%
  \BibitemOpen
  \bibfield  {author} {\bibinfo {author} {\bibfnamefont {D.}~\bibnamefont
  {Jang}}\ and\ \bibinfo {author} {\bibfnamefont {J.~R.}\ \bibnamefont
  {Greer}},\ }\bibfield  {title} {\bibinfo {title} {Transition from a
  strong-yet-brittle to a stronger-and-ductile state by size reduction of
  metallic glasses},\ }\href@noop {} {\bibfield  {journal} {\bibinfo  {journal}
  {Nature Materials}\ }\textbf {\bibinfo {volume} {9}},\ \bibinfo {pages} {215}
  (\bibinfo {year} {2010})}\BibitemShut {NoStop}%
\bibitem [{\citenamefont {Shan}\ \emph {et~al.}(2008)\citenamefont {Shan},
  \citenamefont {Mishra}, \citenamefont {Asif}, \citenamefont {Warren},\ and\
  \citenamefont {Minor}}]{shan2008mechanical}%
  \BibitemOpen
  \bibfield  {author} {\bibinfo {author} {\bibfnamefont {Z.}~\bibnamefont
  {Shan}}, \bibinfo {author} {\bibfnamefont {R.~K.}\ \bibnamefont {Mishra}},
  \bibinfo {author} {\bibfnamefont {S.~S.}\ \bibnamefont {Asif}}, \bibinfo
  {author} {\bibfnamefont {O.~L.}\ \bibnamefont {Warren}},\ and\ \bibinfo
  {author} {\bibfnamefont {A.~M.}\ \bibnamefont {Minor}},\ }\bibfield  {title}
  {\bibinfo {title} {Mechanical annealing and source-limited deformation in
  submicrometre-diameter {Ni} crystals},\ }\href@noop {} {\bibfield  {journal}
  {\bibinfo  {journal} {Nature Materials}\ }\textbf {\bibinfo {volume} {7}},\
  \bibinfo {pages} {115} (\bibinfo {year} {2008})}\BibitemShut {NoStop}%
\bibitem [{\citenamefont {Holian}\ \emph {et~al.}(1991)\citenamefont {Holian},
  \citenamefont {Voter}, \citenamefont {Wagner}, \citenamefont {Ravelo},
  \citenamefont {Chen}, \citenamefont {Hoover}, \citenamefont {Hoover},
  \citenamefont {Hammerberg},\ and\ \citenamefont
  {Dontje}}]{holian1991effects}%
  \BibitemOpen
  \bibfield  {author} {\bibinfo {author} {\bibfnamefont {B.}~\bibnamefont
  {Holian}}, \bibinfo {author} {\bibfnamefont {A.}~\bibnamefont {Voter}},
  \bibinfo {author} {\bibfnamefont {N.}~\bibnamefont {Wagner}}, \bibinfo
  {author} {\bibfnamefont {R.}~\bibnamefont {Ravelo}}, \bibinfo {author}
  {\bibfnamefont {S.}~\bibnamefont {Chen}}, \bibinfo {author} {\bibfnamefont
  {W.~G.}\ \bibnamefont {Hoover}}, \bibinfo {author} {\bibfnamefont
  {C.}~\bibnamefont {Hoover}}, \bibinfo {author} {\bibfnamefont
  {J.}~\bibnamefont {Hammerberg}},\ and\ \bibinfo {author} {\bibfnamefont
  {T.}~\bibnamefont {Dontje}},\ }\bibfield  {title} {\bibinfo {title} {Effects
  of pairwise versus many-body forces on high-stress plastic deformation},\
  }\href@noop {} {\bibfield  {journal} {\bibinfo  {journal} {Physical Review
  A}\ }\textbf {\bibinfo {volume} {43}},\ \bibinfo {pages} {2655} (\bibinfo
  {year} {1991})}\BibitemShut {NoStop}%
\bibitem [{\citenamefont {Baskes}(1999)}]{baskes1999many}%
  \BibitemOpen
  \bibfield  {author} {\bibinfo {author} {\bibfnamefont {M.}~\bibnamefont
  {Baskes}},\ }\bibfield  {title} {\bibinfo {title} {Many-body effects in fcc
  metals: a lennard-jones embedded-atom potential},\ }\href@noop {} {\bibfield
  {journal} {\bibinfo  {journal} {Physical review letters}\ }\textbf {\bibinfo
  {volume} {83}},\ \bibinfo {pages} {2592} (\bibinfo {year}
  {1999})}\BibitemShut {NoStop}%
\bibitem [{\citenamefont {Ziegenhain}\ \emph {et~al.}(2009)\citenamefont
  {Ziegenhain}, \citenamefont {Hartmaier},\ and\ \citenamefont
  {Urbassek}}]{ziegenhain2009pair}%
  \BibitemOpen
  \bibfield  {author} {\bibinfo {author} {\bibfnamefont {G.}~\bibnamefont
  {Ziegenhain}}, \bibinfo {author} {\bibfnamefont {A.}~\bibnamefont
  {Hartmaier}},\ and\ \bibinfo {author} {\bibfnamefont {H.~M.}\ \bibnamefont
  {Urbassek}},\ }\bibfield  {title} {\bibinfo {title} {Pair vs many-body
  potentials: Influence on elastic and plastic behavior in nanoindentation of
  fcc metals},\ }\href@noop {} {\bibfield  {journal} {\bibinfo  {journal}
  {Journal of the Mechanics and Physics of Solids}\ }\textbf {\bibinfo {volume}
  {57}},\ \bibinfo {pages} {1514} (\bibinfo {year} {2009})}\BibitemShut
  {NoStop}%
\bibitem [{\citenamefont {Walsh}(2018)}]{walsh2018rubble}%
  \BibitemOpen
  \bibfield  {author} {\bibinfo {author} {\bibfnamefont {K.~J.}\ \bibnamefont
  {Walsh}},\ }\bibfield  {title} {\bibinfo {title} {Rubble pile asteroids},\
  }\href@noop {} {\bibfield  {journal} {\bibinfo  {journal} {Annual Review of
  Astronomy and Astrophysics}\ }\textbf {\bibinfo {volume} {56}},\ \bibinfo
  {pages} {593} (\bibinfo {year} {2018})}\BibitemShut {NoStop}%
\bibitem [{\citenamefont {Hestroffer}\ \emph {et~al.}(2019)\citenamefont
  {Hestroffer}, \citenamefont {S{\'a}nchez}, \citenamefont {Staron},
  \citenamefont {Bagatin}, \citenamefont {Eggl}, \citenamefont {Losert},
  \citenamefont {Murdoch}, \citenamefont {Opsomer}, \citenamefont {Radjai},
  \citenamefont {Richardson} \emph {et~al.}}]{hestroffer2019small}%
  \BibitemOpen
  \bibfield  {author} {\bibinfo {author} {\bibfnamefont {D.}~\bibnamefont
  {Hestroffer}}, \bibinfo {author} {\bibfnamefont {P.}~\bibnamefont
  {S{\'a}nchez}}, \bibinfo {author} {\bibfnamefont {L.}~\bibnamefont {Staron}},
  \bibinfo {author} {\bibfnamefont {A.~C.}\ \bibnamefont {Bagatin}}, \bibinfo
  {author} {\bibfnamefont {S.}~\bibnamefont {Eggl}}, \bibinfo {author}
  {\bibfnamefont {W.}~\bibnamefont {Losert}}, \bibinfo {author} {\bibfnamefont
  {N.}~\bibnamefont {Murdoch}}, \bibinfo {author} {\bibfnamefont
  {E.}~\bibnamefont {Opsomer}}, \bibinfo {author} {\bibfnamefont
  {F.}~\bibnamefont {Radjai}}, \bibinfo {author} {\bibfnamefont {D.~C.}\
  \bibnamefont {Richardson}}, \emph {et~al.},\ }\bibfield  {title} {\bibinfo
  {title} {Small solar system bodies as granular media},\ }\href@noop {}
  {\bibfield  {journal} {\bibinfo  {journal} {The Astronomy and Astrophysics
  Review}\ }\textbf {\bibinfo {volume} {27}},\ \bibinfo {pages} {1} (\bibinfo
  {year} {2019})}\BibitemShut {NoStop}%
\bibitem [{\citenamefont {Kollmer}\ \emph {et~al.}(2021)\citenamefont
  {Kollmer}, \citenamefont {Featherstone}, \citenamefont {Bullard},
  \citenamefont {Emm}, \citenamefont {Jackson}, \citenamefont {Reid},
  \citenamefont {Shefferman}, \citenamefont {Dove}, \citenamefont {Colwell},\
  and\ \citenamefont {Daniels}}]{kollmer2021probing}%
  \BibitemOpen
  \bibfield  {author} {\bibinfo {author} {\bibfnamefont {J.~E.}\ \bibnamefont
  {Kollmer}}, \bibinfo {author} {\bibfnamefont {J.}~\bibnamefont
  {Featherstone}}, \bibinfo {author} {\bibfnamefont {R.}~\bibnamefont
  {Bullard}}, \bibinfo {author} {\bibfnamefont {T.}~\bibnamefont {Emm}},
  \bibinfo {author} {\bibfnamefont {A.}~\bibnamefont {Jackson}}, \bibinfo
  {author} {\bibfnamefont {R.}~\bibnamefont {Reid}}, \bibinfo {author}
  {\bibfnamefont {S.}~\bibnamefont {Shefferman}}, \bibinfo {author}
  {\bibfnamefont {A.}~\bibnamefont {Dove}}, \bibinfo {author} {\bibfnamefont
  {J.}~\bibnamefont {Colwell}},\ and\ \bibinfo {author} {\bibfnamefont {K.~E.}\
  \bibnamefont {Daniels}},\ }\bibfield  {title} {\bibinfo {title} {Probing
  regolith-covered surfaces in low gravity},\ }in\ \href@noop {} {\emph
  {\bibinfo {booktitle} {EPJ Web of Conferences}}},\ Vol.\ \bibinfo {volume}
  {249}\ (\bibinfo {organization} {EDP Sciences},\ \bibinfo {year} {2021})\ p.\
  \bibinfo {pages} {02005}\BibitemShut {NoStop}%
\bibitem [{\citenamefont {Jacobson}\ and\ \citenamefont
  {Scheeres}(2011)}]{jacobson2011dynamics}%
  \BibitemOpen
  \bibfield  {author} {\bibinfo {author} {\bibfnamefont {S.~A.}\ \bibnamefont
  {Jacobson}}\ and\ \bibinfo {author} {\bibfnamefont {D.~J.}\ \bibnamefont
  {Scheeres}},\ }\bibfield  {title} {\bibinfo {title} {Dynamics of rotationally
  fissioned asteroids: Source of observed small asteroid systems},\ }\href@noop
  {} {\bibfield  {journal} {\bibinfo  {journal} {Icarus}\ }\textbf {\bibinfo
  {volume} {214}},\ \bibinfo {pages} {161} (\bibinfo {year}
  {2011})}\BibitemShut {NoStop}%
\bibitem [{\citenamefont {S{\'a}nchez}\ and\ \citenamefont
  {Scheeres}(2012)}]{sanchez2012simulation}%
  \BibitemOpen
  \bibfield  {author} {\bibinfo {author} {\bibfnamefont {D.~P.}\ \bibnamefont
  {S{\'a}nchez}}\ and\ \bibinfo {author} {\bibfnamefont {D.~J.}\ \bibnamefont
  {Scheeres}},\ }\bibfield  {title} {\bibinfo {title} {DEM simulation of
  rotation-induced reshaping and disruption of rubble-pile asteroids},\
  }\href@noop {} {\bibfield  {journal} {\bibinfo  {journal} {Icarus}\ }\textbf
  {\bibinfo {volume} {218}},\ \bibinfo {pages} {876} (\bibinfo {year}
  {2012})}\BibitemShut {NoStop}%
\bibitem [{\citenamefont {Yasuda}\ \emph {et~al.}(1992)\citenamefont {Yasuda},
  \citenamefont {Torii},\ and\ \citenamefont {Shimizu}}]{yasuda1992self}%
  \BibitemOpen
  \bibfield  {author} {\bibinfo {author} {\bibfnamefont {K.}~\bibnamefont
  {Yasuda}}, \bibinfo {author} {\bibfnamefont {T.}~\bibnamefont {Torii}},\ and\
  \bibinfo {author} {\bibfnamefont {T.}~\bibnamefont {Shimizu}},\ }\bibfield
  {title} {\bibinfo {title} {Self-excited oscillations of a circular disk
  rotating in air},\ }\href@noop {} {\bibfield  {journal} {\bibinfo  {journal}
  {JSME international journal. Ser. 3, Vibration, control engineering,
  engineering for industry}\ }\textbf {\bibinfo {volume} {35}},\ \bibinfo
  {pages} {347} (\bibinfo {year} {1992})}\BibitemShut {NoStop}%
\bibitem [{\citenamefont {Nowinski}(1964)}]{nowinski1964nonlinear}%
  \BibitemOpen
  \bibfield  {author} {\bibinfo {author} {\bibfnamefont {J.~L.}\ \bibnamefont
  {Nowinski}},\ }\bibfield  {title} {\bibinfo {title} {{Nonlinear Transverse
  Vibrations of a Spinning Disk}},\ }\href@noop {} {\bibfield  {journal}
  {\bibinfo  {journal} {Journal of Applied Mechanics}\ }\textbf {\bibinfo
  {volume} {31}},\ \bibinfo {pages} {72} (\bibinfo {year} {1964})}\BibitemShut
  {NoStop}%
\bibitem [{\citenamefont {Mote~Jr}\ \emph {et~al.}(1993)\citenamefont {Mote~Jr}
  \emph {et~al.}}]{mote1993aerodynamically}%
  \BibitemOpen
  \bibfield  {author} {\bibinfo {author} {\bibfnamefont {C.}~\bibnamefont
  {Mote~Jr}} \emph {et~al.},\ }\bibfield  {title} {\bibinfo {title}
  {Aerodynamically excited vibration and flutter of a thin disk rotating at
  supercritical speed},\ }\href@noop {} {\bibfield  {journal} {\bibinfo
  {journal} {Journal of Sound and Vibration}\ }\textbf {\bibinfo {volume}
  {168}},\ \bibinfo {pages} {15} (\bibinfo {year} {1993})}\BibitemShut
  {NoStop}%
\bibitem [{\citenamefont {Renshaw}\ \emph {et~al.}(1994)\citenamefont
  {Renshaw}, \citenamefont {D'Angelo~III},\ and\ \citenamefont
  {Mote~Jr}}]{renshaw1994aerodynamically}%
  \BibitemOpen
  \bibfield  {author} {\bibinfo {author} {\bibfnamefont {A.}~\bibnamefont
  {Renshaw}}, \bibinfo {author} {\bibfnamefont {C.}~\bibnamefont
  {D'Angelo~III}},\ and\ \bibinfo {author} {\bibfnamefont {C.}~\bibnamefont
  {Mote~Jr}},\ }\bibfield  {title} {\bibinfo {title} {Aerodynamically excited
  vibration of a rotating disk},\ }\href@noop {} {\bibfield  {journal}
  {\bibinfo  {journal} {Journal of Sound and Vibration}\ }\textbf {\bibinfo
  {volume} {177}},\ \bibinfo {pages} {577} (\bibinfo {year}
  {1994})}\BibitemShut {NoStop}%
\bibitem [{\citenamefont {Kang}\ and\ \citenamefont
  {Raman}(2006)}]{kang2006vibrations}%
  \BibitemOpen
  \bibfield  {author} {\bibinfo {author} {\bibfnamefont {N.}~\bibnamefont
  {Kang}}\ and\ \bibinfo {author} {\bibfnamefont {A.}~\bibnamefont {Raman}},\
  }\bibfield  {title} {\bibinfo {title} {Vibrations and stability of a flexible
  disk rotating in a gas-filled enclosure—part 2: Experimental study},\
  }\href@noop {} {\bibfield  {journal} {\bibinfo  {journal} {Journal of Sound
  and Vibration}\ }\textbf {\bibinfo {volume} {296}},\ \bibinfo {pages} {676}
  (\bibinfo {year} {2006})}\BibitemShut {NoStop}%
\bibitem [{\citenamefont {Okuizumi}(2007)}]{okuizumi2007equilibrium}%
  \BibitemOpen
  \bibfield  {author} {\bibinfo {author} {\bibfnamefont {N.}~\bibnamefont
  {Okuizumi}},\ }\bibfield  {title} {\bibinfo {title} {Equilibrium of a
  rotating circular membrane under transverse distributed load},\ }\href@noop
  {} {\bibfield  {journal} {\bibinfo  {journal} {Journal of System Design and
  Dynamics}\ }\textbf {\bibinfo {volume} {1}},\ \bibinfo {pages} {85} (\bibinfo
  {year} {2007})}\BibitemShut {NoStop}%
\bibitem [{\citenamefont {Guven}\ \emph {et~al.}(2013)\citenamefont {Guven},
  \citenamefont {Hanna},\ and\ \citenamefont {M{\"u}ller}}]{guven2013whirling}%
  \BibitemOpen
  \bibfield  {author} {\bibinfo {author} {\bibfnamefont {J.}~\bibnamefont
  {Guven}}, \bibinfo {author} {\bibfnamefont {J.}~\bibnamefont {Hanna}},\ and\
  \bibinfo {author} {\bibfnamefont {M.~M.}\ \bibnamefont {M{\"u}ller}},\
  }\bibfield  {title} {\bibinfo {title} {Whirling skirts and rotating cones},\
  }\href@noop {} {\bibfield  {journal} {\bibinfo  {journal} {New Journal of
  Physics}\ }\textbf {\bibinfo {volume} {15}},\ \bibinfo {pages} {113055}
  (\bibinfo {year} {2013})}\BibitemShut {NoStop}%
\bibitem [{\citenamefont {Delapierre}\ \emph {et~al.}(2018)\citenamefont
  {Delapierre}, \citenamefont {Chakraborty}, \citenamefont {Sader},\ and\
  \citenamefont {Pellegrino}}]{delapierre2018wrinkling}%
  \BibitemOpen
  \bibfield  {author} {\bibinfo {author} {\bibfnamefont {M.}~\bibnamefont
  {Delapierre}}, \bibinfo {author} {\bibfnamefont {D.}~\bibnamefont
  {Chakraborty}}, \bibinfo {author} {\bibfnamefont {J.~E.}\ \bibnamefont
  {Sader}},\ and\ \bibinfo {author} {\bibfnamefont {S.}~\bibnamefont
  {Pellegrino}},\ }\bibfield  {title} {\bibinfo {title} {Wrinkling of
  transversely loaded spinning membranes},\ }\href@noop {} {\bibfield
  {journal} {\bibinfo  {journal} {International Journal of Solids and
  Structures}\ }\textbf {\bibinfo {volume} {139}},\ \bibinfo {pages} {163}
  (\bibinfo {year} {2018})}\BibitemShut {NoStop}%
\end{thebibliography}
\end{document}